\documentclass[aps,prb,doublecolumn, amsmath,amssymb,showpacs]{revtex4}
\usepackage{graphicx} 
\usepackage{bm}
\usepackage{dcolumn}
\usepackage{amsmath}
\usepackage{epstopdf}
\usepackage{natbib}
\usepackage{subfig}
\usepackage{comment}

\begin{document}
%
%==========================================================================================
\title{Phase coherence and spectral functions in the two-dimensional excitonic systems}
%==========================================================================================
%
%=======================================================================================================
\author{V. Apinyan\footnote{Institute for Low Temperature and Structure \newline Research, Polish Academy of Sciences\\
PO. Box 1410, 50-950 Wroclaw 2, Poland\newline Tel.: +48 71 343 5021; fax: +48 71 344 1029.\newline \quad\quad 	E-mail address: V.Apinyan@int.pan.wroc.pl.}}

\author{T. K. Kope\'{c}}
%======================================================================================================
 
%=============================================================
\affiliation{%
Institute for Low Temperature and Structure Research, Polish Academy of Sciences\\
PO. Box 1410, 50-950 Wroclaw 2, Poland \\}%
%=============================================================

\date{\today}

%================
\begin{abstract}
%================
%

The nonlocal correlation mechanism between excitonic pairs is considered for a two dimensional exciton system. On the base of the unitary decomposition of the usual electron operator, we include the electron phase degrees of freedom into the problem of interacting excitons. Applying the path integral formalism, we treat the excitonic insulator state (EI) and the Bose-Einstein condensation (BEC) of preformed excitonic pairs as two independent problems. For the BEC of excitons the phase field variables play a crucial role. We derive the expression of the local EI order parameter by integrating out the phase variables. Then, considering the zero temperature limit, we obtain the excitonic BEC transition probability function, by integrating out the fermions.  
We calculate the normal excitonic Green functions for the conduction and valence band electrons and we derive the excitonic spectral functions, both analytically and numerically. Different values of the Coulomb interaction parameter are considered.
\end{abstract}
%Use showkeys class option if keyword
\maketitle

%=======================================
\section{\label{Section_1} Introduction}
%======================================
%
The Coulomb interaction between the electrons and holes gives rise to a very rich physics related to the excitonic systems. The excitons, as composite objects \cite{Moskalenko} with a total zero spin, have a tendency to condense at the very low temperatures, and this is shown for the first time in the sixties of the past millennium. \cite{Keldysh_1,Keldysh_2,Jerome} In general, the Bose-Einstein condensation (BEC) of excitons and the formation of the excitonic insulator (EI) state are considered as the same in the existing literature. \cite{Seki, Zenker_1, Zenker_2, Ihle, Farkasovsky_1, Zenker_3, Batista_1, Batista_2, Czycholl} 
The EI state is a new phase, which develops in the scenario of semimetal (SM)-semiconductor (SC) phase transition, when approaching the transition from the SC side.\cite{Jerome,Halperin} As it is shown in Refs.\onlinecite{Seki, Zenker_1, Zenker_2}, the EI order parameter is non null for a given interval of the Coulomb interaction parameter and for a given value of the valence band hopping amplitude. 
From the interpretation of the results given there, it follows that in the small interaction region, the system is in the Bardeen-Cooper-Schrieffer (BCS) state \cite{Bardeen} with a very weak binding energies of electron-hole pairs, contrary, when approaching from the SC side of the EI state, the system shows BEC behavior with tightly bound excitons,\cite{Ihle,Pethick} thus exhibiting a BCS-BEC type crossover.\cite{Seki, Zenker_1, Zenker_2, Chen} As we mentioned above, in all cited works here, the exciton condensation occurs at the same temperature, as the EI phase transition. It is worth to indicate that the coherence is discussed there in the sense of the direct binding between electrons and holes, without dealing with the phase variables of the quasiparticles. 

However, a series of recent theoretical works suggest the importance of the phase correlations on the phase transition scenario in the excitonic systems. \cite{Snoke_1, Snoke_2, Tomio, Apinyan_1, Apinyan_2,Apinyan_3} Particularly, in Refs.\onlinecite{Snoke_1, Snoke_2}, it is shown theoretically that the EI state and the excitonic BEC are not exactly the same. The importance of the phase coherence in the excitonic pair plasma is discussed there, with a classification of two distinct phase transitions in the excitonic plasma and the discussion about the exciton BEC is provided. It is shown \cite{Snoke_1, Snoke_2, Tomio, Apinyan_2,Apinyan_3} that in the low density limit of the excitonic pairs, the critical temperature of excitonic BEC should be much smaller than the temperature of the pair formation.

In the high e-h density limit we have the convergence of theories, since in this case the transition lines of excitonic condensation and of that of the pair formation are coinciding.  Indeed, where the mean distance between the particles is shorter than the excitonic Bohr radius, the weakly bound e-h pairs behave like the Cooper pairs in the conventional superconductors at sufficiently low temperatures. \cite{Keldysh_2, Jerome, Micnas} In this case, the condensation is of the BCS type. In Ref.\ \onlinecite{Tomio}, the authors employ the two-band Hubbard model within the self-consistent $t$-matrix approximation to show that in the low density limit the gas of free excitons undergoes the BEC phase transition at the very low temperatures, and the BEC temperature transition line is not coinciding with that of the pair formation. In fact, the BEC of the excitonic pairs is possible only when the macroscopic phase coherence is present in the system.\cite{Snoke_1} 
The EI state is an excitonium state, where the incoherent e-h bound pairs are formed and furthermore, at the lower temperatures, the BEC of excitons appears in consequence of reconfiguration and coherent condensation of preformed excitonic pairs.

In the weak-coupling limit, the transition to e-h condensed phase is related to the relative motion between electrons and holes,\cite{Tomio} which implies the BCS-like regime and is in contrast to the case of strong-coupling, when the BEC state is related to the motion of the center of mass of excitons. The e-h mass difference in the BCS-BEC transition scenario leads to a large suppression of the BEC transition temperature, which is proved to not be the same as the excitonic pair formation temperature. \cite{Tomio, Apinyan_2,Apinyan_3} This is in contrast with the previous treatments,\cite{Seki,Zenker_1,Zenker_2,Ihle,Farkasovsky_1, Zenker_3,Batista_1,Batista_2,Czycholl} where the 
EI state is associated with the BEC state of excitons, as to be identical. 
We treat the e-h system in the frames of (spinless) two-dimensional (2D) extended Falicov-Kimball model (EFKM), as a purposeful model, to include the $f$-$f$ hopping mechanism that could be also responsible for the exciton formation.\cite{Farkasovsky_1} Using the electron operator representation, we address the role of the phase factor in the context of the interacting excitons. As a first step of the theory, we obtain the EI order parameter by employing the functional integration technique and we discuss the stability region (in the $T$-$U$ plan) of the exciton pair formation. Furthermore, at the zero temperature limit, we integrate out the fermions and we discuss the obtained phase action and the phase stiffness. We show that the phase stiffness in the system is directly related to the exciton condensation in the 2D excitonic system at zero temperature. We calculate the phase stiffness parameter for different values of the $f$-band hopping amplitude. 

Then, turning to the phase sector, we employ the Bogoliubov mean field displacement approximation, for the bosonic charge variables and, hence, we separate the excitonic condensate part in the e-h paired plasma (excitonium).
Furthermore, we calculate the exciton BEC transition probability, as a function of the attractive Coulomb interaction parameter $U/t$, which is normalized to the hopping integral  of the conduction band electrons. By using the Fourier space representation, we give the expressions of the total normal Green functions for the $f$ and $c$-band electrons and we emphasize on the phase dependence of those functions.
 As the consequence, we obtain the frequency dependent normal spectral functions, at the zero temperature case and, furthermore, the phase-coherent density of states (DOS).
The numerical evaluations of normal DOS functions, for the $f$ and $c$ band electrons, show a gapless character of the spectrum of excitations, in contrast to the traditionally admitted incoherent DOS behavior. We show that the hybridization-gap is totally absent for all frequency modes and for all values of the Coulomb interaction parameter. We argue that the gapless behavior in the DOS spectra is a result of competition of two independent excitations in the system: the phase fluctuations and strong quantum coherence effects at zero temperature limit. Note, that a similar gapless character in the DOS spectrum of cold excitons is observed recently in Ref.\onlinecite{Lemanskii}, where this effect is associated with metallic charge-density-wave phase and it is driven by the strong electron correlations.

The paper is organized as follows. In Sec. \ref{sec:Section_2}, we introduce the model Hamiltonian. The electron factorization and resulting phase action are presented in the Section \ref{sec:Section_3}.
In the Sec. \ref{sec:Section_4}, we get the effective fermionic action for the EI state in the system. The
numerical results are presented
there. In the Sec. \ref{sec:Section_5} we integrate out the fermions and we obtain the phase stiffness parameter, both analytically and numerically. In the Sec. \ref{sec:Section_6} we discuss the 2D excitonic BEC at $T=0$ and we calculate the ecxitonic BEC transition probability function. The Sec. \ref{sec:Section_7} is devoted to the calculation of the single particle spectral functions and density of states. At the end of the Sec. \ref{sec:Section_7} we give the numerical evaluations for DOS functions and we discuss the obtained results.
Meanwhile, an experimental technique is proposed to prove directly the DOS behavior. Finally, in the Sec. \ref{sec:Section_8} we give a conclusion of our results. The theoretical calculation of the phase action is given in the Appendix.
%
%===================================================
\section{\label{sec:Section_2} The method}
%===================================================
%
%===================================================
\subsection{\label{sec:Section_2_1} EFKM Hamiltonian}
%===================================================
%
The Hamiltonian of the spinless EFKM model is given by
\begin{eqnarray}
{\cal{H}}=-\sum_{\rm x=f,c}\left\{ \sum_{\left\langle {\bf{r}},{\bf{r}}'\right\rangle}  t_{\rm x}\left[\bar{\rm x}({{\bf{r}}}){\rm x}({{\bf{r}}}')+h.c.\right]+\left(\mu_{\rm x}-\epsilon_{\rm x}\right)\sum_{{\bf{r}}}n_{\rm x}({\bf{r}})-\frac{U}{2}\sum_{{\bf{r}}}n_{\rm x}({{\bf{r}}})n_{\tilde{\rm x}}({{\bf{r}}})\right\}.
\label{Equation_1}
\end{eqnarray}
Here, the operator $\bar{\rm x}({{\bf{r}}})$ (${\rm x}({{\bf{r}}})$) creates (annihilates) an $f$ or $c$ electron at the lattice position ${\bf{r}}$, the notation $\tilde{\rm x}$ in the last term in Eq.(\ref{Equation_1}) means the orbital opposite to $\rm x$, the summation $\left\langle {\bf{r}}, {\bf{r}}' \right\rangle$ runs over pairs of nearest neighbor (n.n.) sites on the 2D square lattice. The spin degrees of freedom have been ignored for simplicity. Next, $t_{x}$ is the hopping amplitude for $\rm x$-electrons and $\epsilon_{\rm x}$ is the corresponding on-site energy level. 
The sign of the product $t_{\rm x}t_{\tilde{\rm x}}$ determines the type of semiconductor, for $t_{\rm x}t_{\tilde{\rm x}}<0$ ($t_{\rm x}t_{\tilde{\rm x}}>0$) we have the direct (indirect) band gap semiconductor. The case $t_{f}\equiv 0$ corresponds to that of the dispersionless $f$ band and usual Falicov-Kimball model\cite{Falicov} (FKM) could be derived (in this case, the local $f$-electron number is conserved).

The on-site (local) interaction parameter $U$, in the last term of the Hamiltonian in Eq.(\ref{Equation_1}), is the Coulomb repulsion parameter (interorbital) between the electrons in the $f$ and $c$ orbitals. As we will see later on, the strength of the local Coulomb interaction will tune the SM-SC transition in the system and the formation of the local EI state in the excitonic system. In the case of the degenerated $f$ and $c$ bands, i.e. when $\epsilon_{\rm x}=\epsilon_{\tilde{\rm x}}$ and $t_{\rm x}=t_{\tilde{\rm x}}$, the EFKM model reduces to the standard Hubbard model.\cite{Hubbard}
Furthermore, we adjust the chemical potentials $\mu_{\rm x}$ and $\mu_{\tilde{\rm x}}$ in order to maintain separate the number of electrons in $f$ and $c$ orbitals. Then, the equilibrium value of chemical potential $\mu\equiv \mu_{\rm x}=\mu_{\tilde{\rm x}}$ in Eq.(\ref{Equation_1}) will be determined from the half-filling condition, i.e. we suppose that $\left\langle n_{\rm x}({\bf{r}})\right\rangle+\left\langle n_{\tilde{\rm x}}({\bf{r}})\right\rangle=1$. In what follows, we assume a band structure with a direct band gap, i.e. $t_{\rm x}t_{\tilde{\rm x}}<0$ and without the loss of generality the $c$ electrons are considered to be `` light'', while the $f$ electrons are ``heavy'', i.e. $t_{f}<1$, and the hopping integral for $c$ electrons is taken to be the unit of the energy scale $t_{c}=1$. Throughout the paper, we set $k_{B} = 1$ and $\hbar=1$, and, the lattice
constant, $d = 1$. For frequency notations, we keep the symbol $\nu$ for fermions and $\omega$ - for bosons. We set also $\epsilon_{c}=0$.

The genuine feature of the EFKM Hamiltonian in Eq.(\ref{Equation_1}) is that it is equivalent to the asymmetric Hubbard model, if we associate for orbitals $c$ and $f$ the spin variables, thus replacing the fermionic Hilbert space with the pseudo-fermionic one, and then by linearizing the interaction term via the bosonic states (see in Ref.\ \onlinecite{Zenker_1}).
%
%==============================================================================================
\subsection{\label{sec:Section_2_2} Hubbard-Stratanovich linearisation}
%==============================================================================================
%
It is more convenient to write the EFKM Hamiltonian given in Eq.(\ref{Equation_1}) in more symmetric form, suitable for the mean-field decoupling. 
\begin{eqnarray}
{\cal{H}}=-\sum_{\left\langle {\bf{r}},{\bf{r}}'\right\rangle \atop {\rm x}=f,c }t_{\rm x}\left[\bar{\rm x}({{\bf{r}}}){\rm }{\rm x}({{\bf{r}}}')+h.c.\right]-\bar{\mu}\sum_{{\bf{r}}}n({\bf{r}})+\frac{\epsilon_{c}-\epsilon_{f}}{2}\sum_{{\bf{r}}}\tilde{n}({\bf{r}})
+\frac{U}{4}\sum_{{\bf{r}}}\left[n^{2}({\bf{r}})-\tilde{n}^{2}({\bf{r}})\right].
\label{Equation_2}
\end{eqnarray}
The chemical potential $\bar{\mu}$ is $\bar{\mu}=\mu-\bar{\epsilon}$, where $\bar{\epsilon}=\left(\epsilon_{c}+\epsilon_{f}\right)/2$. The short hand notations were introduced in Eq.(\ref{Equation_2}): $n({\bf{r}})=n_{c}({\bf{r}})+n_{f}({\bf{r}})$ and $\tilde{n}({\bf{r}})=n_{c}({\bf{r}})-n_{f}({\bf{r}})$ in order to simplify the calculations.

\begin{figure}
\begin{center}z
\includegraphics[width=200px,height=190px]{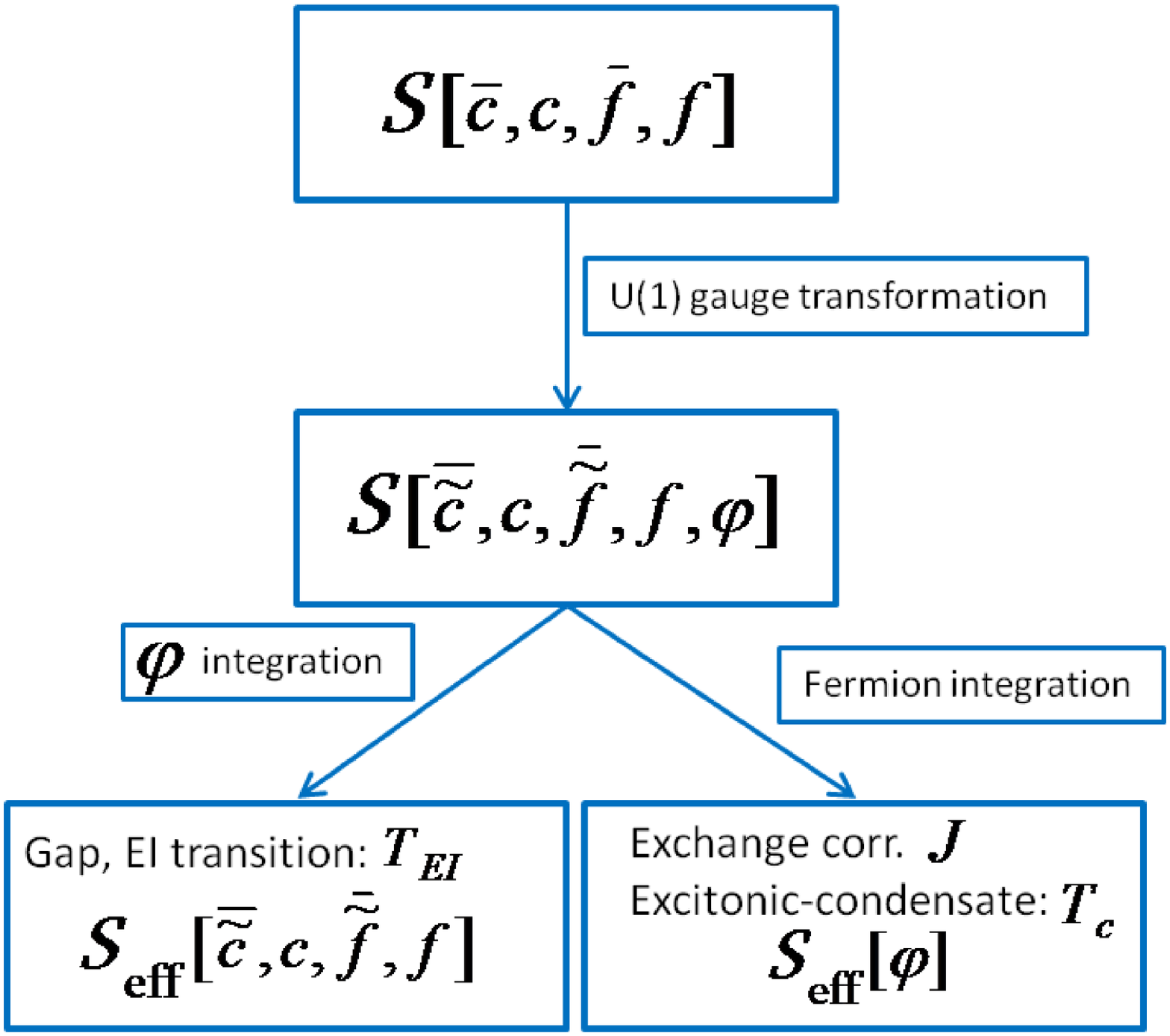}% Here is how to import EPS art
\caption{\label{fig:Fig_1}(Color online) The functional integration procedure.}
\end{center}
\end{figure}

The dealing with fermions within the path integral method, requires introduction of the Grassmann variables ${c}({{\bf{r}}}\tau)$ and ${f}({{\bf{r}}}\tau)$ at each site ${\bf{r}}$ and at each imaginary time $\tau$. The latest varies in the interval $0\leq \tau \leq\beta$, where $\beta=1/T$ (with $T$ being the thermodynamic temperature). The variables ${c}({{\bf{r}}}\tau)$ and ${f}({{\bf{r}}}\tau)$ satisfy the anti-periodic boundary conditions ${\rm x}({{\bf{r}}}\tau)=-{\rm x}({{\bf{r}}}\tau+\beta)$. 
The partition function of the system of the fermions, written as a functional integral over the Grassmann field, is
\begin{eqnarray}
{\cal{Z}}=\int\left[{\cal{D}}\bar{f}{\cal{D}}f\right]\int \left[{\cal{D}}\bar{c}{\cal{D}}c\right]e^{-{\cal{S}}[\bar{c},c, \bar{f},f]},
\label{Equation_3}
\end{eqnarray} 
where the action in the expression of the exponent is given as
\begin{eqnarray}
{\cal{S}}[\bar{c},c, \bar{f},f]=\sum_{{\rm x}=f,c}{\cal{S}}_{B}[\bar{\rm x},{\rm x}]+\int^{\beta}_{0}d\tau {\cal{H}}(\tau).
\label{Equation_4}
\end{eqnarray} 
Here ${\cal{S}}_{B}[\bar{\rm x},{\rm x}]$ is the fermionic Berry term for the $f$ and $c$ -band electrons. It is defined as
\begin{eqnarray}
{\cal{S}}_{B}[\bar{\rm x},{\rm x}]=\sum_{{\bf{r}}}\int^{\beta}_{0}d\tau \bar{\rm x}({\bf{r}}\tau)\frac{\partial}{\partial{\tau}}{\rm x}({\bf{r}}\tau).
\label{Equation_5}
\end{eqnarray}
Next, we decouple quadratic density terms in Eq.(\ref{Equation_2}) using the Hubbard-Stratonovich (HS) transformation.\cite{Negele} We do not present here the calculation details,\cite{Apinyan_2,Apinyan_3} but just the final result for the total action. It reads as
\begin{eqnarray}
&&{\cal{S}}[\bar{c},c,{\bar{f}},f,\varphi]={{S}}_{\rm eff}\left[\varphi\right]+\sum_{{\rm x}=f,c}{\cal{S}}_{B}[\bar{\rm x},{\rm x}]-\sum_{\left\langle{\bf{r}},{\bf{r}}' \right\rangle \atop x=f,c}\int^{\beta}_{0}d\tau t_{\rm x}\left[\bar{\rm x}({{\bf{r}}}\tau){\rm x}({{\bf{r}}}'\tau)+h.c.\right]
+\sum_{{\bf{r}}}\int^{\beta}_{0}d\tau \left[{\mu}_{n}n({\bf{r}}\tau)+\mu_{\tilde{n}}\tilde{n}({\bf{r}}\tau)\right].
\label{Equation_6}
\end{eqnarray}
After the HS linearisation, we got the
total action of the system that is linear in terms of
fermion density operators $n({\bf{r}}\tau)$ and $\tilde{n}({\bf{r}}\tau)$.
Here, we see that the additional phase variables $\varphi$, and the phase action ${\cal{S}}_{\rm eff}\left[\varphi\right]$ are present. \cite{Apinyan_1,Apinyan_2,Apinyan_3} This is due to the fact that the time derivative of the phase variables is equal to the periodic part of decoupling field coupled to the total density function $n({\bf{r}},\tau)$ (for the details see in Refs.\onlinecite{Apinyan_1,Apinyan_2,Apinyan_3}). The effective chemical potentials $\mu_{n}$ and $\mu_{\tilde{n}}$, appearing after the HS decoupling procedure and saddle-point analysis, are 
\begin{eqnarray}
&&{\mu}_{n}=\frac{Un}{2}-\bar{\mu},
\label{Equation_7}
\newline\\
&&\mu_{\tilde{n}}=\frac{\epsilon_{c}-\epsilon_{f}}{2}-\frac{U\tilde{n}}{2}.
\ \ \ 
\label{Equation_8}
\end{eqnarray}
The phase action ${{S}}_{\rm eff}[\varphi]$ is given by\cite{Apinyan_2,Apinyan_3} 
\begin{eqnarray}
{{S}}_{\rm eff}[\varphi]=\sum_{{\bf{r}}}\int^{\beta}_{0}d\tau\left[\frac{\dot{\varphi}^{2}({\bf{r}}\tau)}{U}-\frac{2\bar{\mu}}{iU}\dot{\varphi}({\bf{r}}\tau)-i\dot{\varphi}({\bf{r}}\tau)n({\bf{r}}\tau)\frac{}{}\right].
\label{Equation_9}
\end{eqnarray}
Thus, the introduction of phase variables into the problem divides the system into two separate parts. One, related to the bosonic phase sector with the phase variables $\varphi$ and another one, the typical fermionic part. 
The partition function of the system is 
\begin{eqnarray}
{\cal{Z}}=\int \left[{\cal{D}}\bar{f}{\cal{D}}f\right]\int\left[{\cal{D}}\bar{c}{\cal{D}}c\right]\int\left[{\cal{D}}\varphi\right]e^{-{\cal{S}}[\bar{c},c,{\bar{f}},f,\varphi]}.
\label{Equation_10}
\end{eqnarray}
The action, in the form given in Eq.(\ref{Equation_6}), is now suitable for derivation of the effective phase action, and of the fermionic action (see in Fig.~\ref{fig:Fig_1}, for the general integration procedure). 
By performing the integration over the phase field in Eq.(\ref{Equation_10}), we should take into account, that the bosonic phase field configurations satisfy the boundary conditions
\begin{eqnarray}
\varphi({\bf{r}}\beta)-\varphi({\bf{r}}0)=2\pi{m({\bf{r}})},
\label{Equation_11}
\end{eqnarray}
where the winding numbers $m\left({\bf{r}}\right)$ characterize all paths in the configuration space. Hence, any two paths, which have different winding numbers, cannot be continuously  transformed from one into another, and in order to include all the possible phase path contributions, we have to sum
over all topologically inequivalent phase configurations, described by their winding numbers
\begin{eqnarray}
\int\left[{\cal{D}}\varphi\right]...\equiv \sum_{\left\{m({\bf{r}})\right\}}\prod_{{\bf{r}}}\int^{2\pi}_{0}d\varphi_{0}({\bf{r}}) \prod_{{\bf{r}}}\int^{\varphi({\bf{r}}\beta)=\varphi_{0}\left({\bf{r}}\right)+2{\pi}m({\bf{r}})}_{\varphi\left({\bf{r}}0\right)=\varphi_{0}\left({\bf{r}}\right)}d\varphi({\bf{r}}\tau)... \;. 
\label{Equation_12}
\end{eqnarray}
Thus, the integration over the phase field amounts the integration over the $\beta$-periodic field $\varphi({\bf{r}}\tau)$ and the summation over a set of U(1) winding numbers $m({\bf{r}})$.
%
%======================================================================================
%======================================================================================
%
%===================================================================
\section{\label{sec:Section_3} The U(1) transformation}
\begin{figure}[htb]
\centering
  \subfloat[\label{fig:Fig_2}The gauge representation of exciton. The dashed arrows represent the direction of bosonic phase-flux around fermionic particles. The solid arrows represent the spin of the electron or hole.]{%
   \centerline{\includegraphics[width=110px,height=110px]{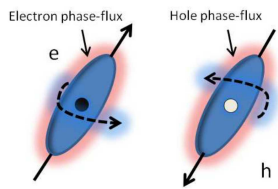}}}\hfill 
  \subfloat[ \label{fig:Fig_3} {The gauge representation of the Cooper pair. The dashed arrows represent the direction of the bosonic phase-flux around fermionic particles. The solid arrows represent the spins of the electrons.}]{%
    \centerline{\includegraphics[width=110px,height=110px]{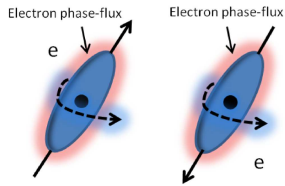}}}
    \caption{The gauge representation of the exciton and Cooper pair.}
\end{figure}

%===================================================================
%
Next, we perform the local gauge transformation of the fermionic Grassmann variables $f({\bf{r}}\tau)$ and $c({\bf{r}}\tau)$ in
order to separate the different gauge degrees of freedom. This procedure will automatically
eliminate also the last imaginary term appearing in the
expression of phase action in Eq.(\ref{Equation_9}). For 
the electrons of the $f$ band, the U$(1)$ transformation is given by 
\begin{eqnarray}\left[
\begin{array}{cc}
f({\bf{r}}\tau) \\
\bar{f}({\bf{r}}\tau)
\end{array}
\right]=\left[
\begin{array}{cc}
e^{i\varphi({\bf{r}}\tau)}&0\\
0& e^{-i\varphi({\bf{r}}\tau)}
\end{array}
\right] \cdot
\left[
\begin{array}{cc}
b({\bf{r}}\tau) \\
\bar{b}({\bf{r}}\tau)
\end{array}
\right].
\label{Equation_13}
\end{eqnarray}
Here, we used the bosonic phase variables $\varphi({\bf{r}}\tau)$, introduced in Eq.(\ref{Equation_6}). For the $c$-orbital electrons the similar transformation is
\begin{eqnarray}\left[
\begin{array}{cc}
c({\bf{r}}\tau) \\
\bar{c}({\bf{r}}\tau)
\end{array}
\right]=\left[
\begin{array}{cc}
e^{i\varphi({\bf{r}}\tau)}&0\\
0& e^{-i\varphi({\bf{r}}\tau)}
\end{array}
\right] \cdot\left[
\begin{array}{cc}
a({\bf{r}}\tau) \\
\bar{a}({\bf{r}}\tau)
\end{array}
\right].
\label{Equation_14}
\end{eqnarray}
As a result the electron appears in the theory like a composite object, in the form of the fermion particle with the attached U(1)
phase ``flux-tube'' (we presented it schematically in Figs.~\ref{fig:Fig_2} and ~\ref{fig:Fig_3}
like the bosonic clouds (in blue), surrounding fermionic
particles). The flows of phase flux (see dashed arrows
in Figs.~\ref{fig:Fig_2} and ~\ref{fig:Fig_3}), are independent of spin of the fermions, but
depend on the particle type. For the electrons and holes, the superflow has the opposite direction, thus, no supercurrent could traverse across a system (see in Fig.~\ref{fig:Fig_2},
no Meissner effect \cite{Bardeen} happens in this case). This is in
contrast to the usual BCS picture \cite{Bardeen} (presented in Fig.~\ref{fig:Fig_3}), where the pairing is between electrons and the superflow of the phase flux tubes has the same direction, leading to the appearance of supercurrent in the system.
%
%==============================================================================
%==============================================================================

In fact, the electron factorization in terms of two variables has an unprecedented impact on the whole theory. As it is shown in Refs.\onlinecite{Apinyan_1,Apinyan_2,Apinyan_3}, this procedure could explain the excitonic Berezinskii-Kosterlitz-Thouless phase transition in the 2D excitonic system \cite{Apinyan_1} and, also, describes the excitonic BEC transition for a three-dimensional system of excitons. \cite{Apinyan_2}

After the transformation procedure, the action in Eq.(\ref{Equation_6}) takes the following form 
\begin{eqnarray}
{\cal{S}}[\bar{a},a,{\bar{b}}, b,\varphi]&&={\cal{S}}_{0}[\varphi]+\sum_{{\rm x}=b,a}{\cal{S}}_{B}[\bar{\rm x},{\rm x}]-\sum_{\langle{\bf{r}},{\bf{r}}'\rangle \atop {\rm x}=b,a}\int^{\beta}_{0}d\tau t_{\rm x}\left[\bar{\rm x}({{\bf{r}}}\tau){\rm x}({{\bf{r}}}'\tau)e^{-i\left[\varphi({{\bf{r}}}\tau)-\varphi({{\bf{r}}}'\tau)\right]}+h.c.\right]
\nonumber\\
&&+\sum_{{\bf{r}}}\int^{\beta}_{0}d\tau \left[{\mu}_{n}n({\bf{r}}\tau)+\mu_{\tilde{n}}\tilde{n}({\bf{r}}\tau)\right],
\label{Equation_15}
\end{eqnarray}
with the new phase action ${\cal{S}}_{0}[\varphi]$, which doesn't contains the fluctuating imaginary term (see the last term in the action in Eq.(\ref{Equation_9}))
\begin{eqnarray} {\cal{S}}_{0}[\varphi]=\sum_{{\bf{r}}}\int^{\beta}_{0}d\tau\left[\frac{\dot{\varphi}^{2}({\bf{r}}\tau)}{U}-\frac{2\bar{\mu}}{iU}\dot{\varphi}({\bf{r}}\tau)\right].
\label{Equation_16}
\end{eqnarray}
For the next, we will put $t_{b}\equiv\tilde{t}$ and $t_{c}\equiv t$.
Then, the partition function of the system in new variables is
\begin{eqnarray}
{\cal{Z}}=\int\left[{\cal{D}}\bar{b}{\cal{D}}b\right]\int\left[{\cal{D}}\bar{a}{\cal{D}}a\right]\int\left[{\cal{D}}\varphi\right] e^{-{\cal{S}}[\bar{a},a,{\bar{b}}, b,\varphi]}.
\label{Equation_17}
\end{eqnarray}
This form of the partition function will be the starting point for deriving the effective actions for the fermions and for the phase sector (see in Fig.~\ref{fig:Fig_1}). 
%
%
%=====================================================
\section{\label{sec:Section_4} Local correlations and excitonic gap }
%=====================================================
%
%=====================================================
\subsection{\label{sec:Section_4_1} EI state }
\begin{figure}
\begin{center}
\includegraphics[width=180px,height=170px]{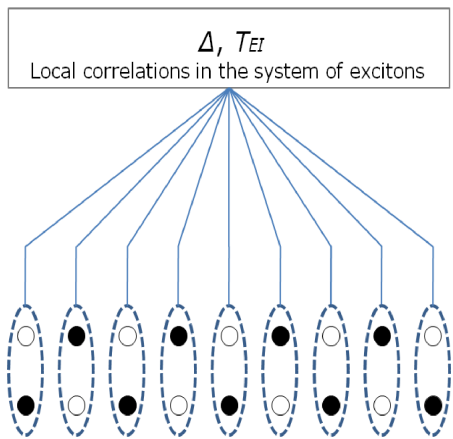}% Here is how to import EPS art
\caption{\label{fig:Fig_4}(Color online) The local correlations, mediated by the Coulomb attraction $U$ in the excitonic system, forming the EI state at the temperature $T_{EI}$.}
\end{center}
\end{figure}

%=====================================================
%
In this section, we will show how the local correlations between the electrons and holes lead to the insulator phase in the system called the ``excitonic insulator''.\cite{Moskalenko}
The EI low-temperature phase is characterized by a local excitonic order parameter (excitonic gap).
The non-vanishing of the expectation value
\begin{eqnarray}
\Delta=U\left\langle \bar{a}({{\bf{r}}}\tau)b({{\bf{r}}}\tau)\right\rangle
\label{Equation_18}
\end{eqnarray}
signals the appearance of the electron-hole bound pairs, which manifests as a gap in the excitation spectrum and signals the presence of the EI state. The EI state develops from the local, on-site electron-hole correlations (see in Fig.~\ref{fig:Fig_4}).
%
%======================================================================================
%======================================================================================
%
Furthermore, by applying the standard Hartree-Fock approximation to the interaction term in the initial Hamiltonian in Eq.(\ref{Equation_1}), we will decouple it. The procedure is described in details in Refs.\onlinecite{Seki, Czycholl, Apinyan_1,Apinyan_2,Apinyan_3}. Then, following the picture presented
in Fig.~\ref{fig:Fig_1}, we integrate out the phase variables in the partition function given in Eq.(\ref{Equation_17}) to obtain the effective fermionic action of the model
\begin{eqnarray}
{\cal{Z}}=\int\left[D\bar{b}Db\right]\left[D\bar{a}Da\right] e^{-{\cal{S}}_{\cal{F}}\left[\bar{a},a,\bar{b},b\right]},
\nonumber\\
\ \ \ 
\label{Equation_19}
\end{eqnarray}
where the effective action, in the expression of the exponent, is given by the relation
\begin{eqnarray}
{\cal{S}}_{\cal{F}}\left[\bar{a},a,\bar{b},b\right]=-\ln\int\left[D\varphi\right] e^{-{S}\left[\bar{a},a,\bar{b},b,\varphi\right]}.
\nonumber\\
\ \ \ 
\label{Equation_20}
\end{eqnarray}
The Fourier transformations of fermionic variables $b({\bf{r}}\tau)$ and $a({\bf{r}}\tau)$ are given by 
\begin{eqnarray}
{\rm x}({\bf{r}}\tau)=\frac{1}{{\beta{N}}}\sum_{{\bf{k}}\nu_{n}}{\rm x}_{{\bf{k}}}(\nu_{n})e^{i{\bf{k}}\cdot{\bf{r}}-i\nu_{n}\tau}
\label{Equation_21}
\end{eqnarray}
with ${\rm x}=b,a$. The number $N$, in Eq.(\ref{Equation_21}), is the number of total lattice sites for an orbital, and $\nu_{n}={\pi(2n+1)}/{\beta}$ are the Fermi-Matsubara frequencies\cite{Abrikosov} with $n=0,\pm1,\pm2,...$. 

The effective fermionic action ${\cal{S}}_{\cal{F}}\left[\bar{a},a,\bar{b},b\right]$, obtained after phase integration, is (see in Fig.~\ref{fig:Fig_1}, in the Section \ref{sec:Section_3})  
\begin{eqnarray}
{\cal{S}}_{\cal{F}}\left[\bar{a},a,\bar{b},b\right]=&&\frac{1}{\beta{N}}\sum_{{\bf{k}},\nu_{n}}\bar{a}_{{\bf{k}}}(\nu_{n})\left(\bar{\epsilon}_{a}-i\nu_{n}-{t}_{{\bf{k}}}\right)a_{{\bf{k}}}(\nu_{n})+\frac{1}{\beta{N}}\sum_{{\bf{k}},\nu_{n}}\bar{b}_{{\bf{k}}}(\nu_{n})\left(\bar{\epsilon}_{b}-i\nu_{n}-\tilde{t}_{{\bf{k}}}\right)\tilde{f}_{{\bf{k}}}(\nu_{n})
\nonumber\\
&&-\frac{\bar{\Delta}}{\beta{N}}\sum_{{\bf{k}},\nu_{n}}\bar{a}_{{\bf{k}}}(\nu_{n})b_{{\bf{k}}}(\nu_{n})
-\frac{{\Delta}}{\beta{N}}\sum_{{\bf{k}},\nu_{n}}\bar{b}_{{\bf{k}}}(\nu_{n})a_{{\bf{k}}}(\nu_{n}).
\label{Equation_22}
\end{eqnarray}
We have obtained in Eq.(\ref{Equation_22}) the Hartree quasiparticle-energies $\bar{\epsilon}_{b}$ and $\bar{\epsilon}_{a}$ as 
\begin{eqnarray}
\bar{\epsilon}_{b}=\epsilon_{f}-\mu+Un_{a}+i\left\langle\dot{\varphi}({{\bf{r}}}\tau)\right\rangle,
\label{Equation_23}
\newline\\
\bar{\epsilon}_{a}=\epsilon_{c}-\mu+Un_{b}+i\left\langle\dot{\varphi}({{\bf{r}}}\tau)\right\rangle.
\label{Equation_24}
\end{eqnarray}
Note, that we kept naturally the same notations for quasiparticle band energy parameters $\epsilon_{c}$ and $\epsilon_{f}$. Furthermore, $n_{b}$ and $n_{a}$ in Eqs.(\ref{Equation_23}) and ({\ref{Equation_24}}) are $b$ and $a$-fermion average densities $n_{{\rm x}}=\left\langle n_{\rm x}({\bf{r}}\tau)\right\rangle=\left\langle \bar{\rm x}({\bf{r}}\tau){\rm x}({\bf{r}}\tau)\right\rangle$. In addition, the usual Hartree shifts are given by the terms $Un_{\rm x}$ in Eqs.(\ref{Equation_23}) and (\ref{Equation_24}). The last imaginary terms in Eqs.(\ref{Equation_23}) and ({\ref{Equation_24}}) in the form $i\left\langle\dot{\varphi}({{\bf{r}}}\tau)\right\rangle$ are completely unimportant for the case of the homogeneous phase distributions in different bands, which we consider here. These terms are related to the phase fluctuation decay in the case of the inhomogeneous phase distribution, when an external electric of magnetic field (in the quantum Hall regime) are applied to the system, causing the finite lifetime of the quasiparticles on different energy band levels.

Next, ${t}_{{\bf{k}}}$ and $\tilde{t}_{{\bf{k}}}$ in Eq.(\ref{Equation_22}) are band-renormalized hopping amplitudes ${t}_{{\bf{k}}}=2t{\mathrm{g}}_{B}\epsilon\left({{\bf{k}}}\right)$ and $\tilde{t}_{{\bf{k}}}=2\tilde{t}{\mathrm{g}}_{B}\epsilon\left({{\bf{k}}}\right)$, where ${\mathrm{g}}_{B}$ is the bandwidth renormalization factor 
\begin{eqnarray}
\mathrm{g}_{B}=\left.\left\langle e^{-i[\varphi({{\bf{r}}}\tau)-\varphi({{\bf{r}}}'\tau)]} \right\rangle\right|_{|{\bf{r}}-{\bf{r}}'|={{d}}}
\label{Equation_25}
\end{eqnarray}
and $\epsilon\left({{\bf{k}}}\right)$ is the 2D lattice dispersion 
\begin{eqnarray}
\epsilon\left({{\bf{k}}}\right)=\cos(k_{x}d_{x})+\cos(k_{y}d_{y}), 
\label{Equation_26} 
\end{eqnarray}
with $d_{\alpha}$ ($\alpha=x,y$), being the components of the lattice spacing vector ${\bf{d}}={\bf{r}}-{\bf{r}}'$ with ${\bf{r}}$ and ${\bf{r}}'$ n.n. site positions. For the simple square-plane, we have $d_{\alpha}\equiv d$.  
The calculation of ${\mathrm{g}_{B}}$ within the self-consistent harmonic approximation \cite{Simanek, Wood_Stroud, Chakravarty,Kleinert,Seunghwan} is discussed in the Ref.\onlinecite{Apinyan_1}. The evaluation of the factor $g_{B}\left({\bf{r}}-{\bf{r}}'\right)$ shows that, at $T=0$, it is equal identically to $1$.

Next, we define the inverse Green function matrix 
\begin{eqnarray}
\hat{\cal{G}}^{-1}({\bf{k}}\nu_{n})=
\left[
\begin{array}{cc}
{\cal{E}}^{a}_{{\bf{k}}}(\nu_{n})
 & -\bar{\Delta}  \\
-\Delta & {\cal{E}}^{b}_{{\bf{k}}}(\nu_{n})
\end{array}
\right],
\label{Equation_27}
\end{eqnarray}
and we rewrite the effective fermionic action in Eq.(\ref{Equation_22}) in more compact Nambu form
\begin{eqnarray}
{\cal{S}}_{\cal{F}}\left[\bar{a},a,\bar{b},b\right]=\frac{1}{\beta{N}}\sum_{{\bf{k}},\nu_{n}}\left[\bar{a}_{\bf{k}}(\nu_{n}),\bar{b}_{\bf{k}}({\nu_{n}})\right]\hat{\cal{G}}^{-1}({\bf{k}}\nu_{n})\left[\begin{array}{cc}
{a}_{\bf{k}}(\nu_{n})\\
{b}_{\bf{k}}(\nu_{n})
\end{array}
\right],
\label{Equation_28}
\end{eqnarray}
where the single-particle Bogoliubov quasienergies ${E}^{b}_{{\bf{k}}}(\nu_{n})$ and ${E}^{a}_{{\bf{k}}}(\nu_{n})$ are given after Eqs.(\ref{Equation_23}) and (\ref{Equation_24}) as
\begin{eqnarray}
{E}^{b}_{{\bf{k}}}(\nu_{n})=\bar{\epsilon}_{b}-i\nu_{n}-\tilde{t}_{{\bf{k}}},
\label{Equation_29}
\newline\\
{E}^{a}_{{\bf{k}}}(\nu_{n})=\bar{\epsilon}_{a}-i\nu_{n}-{t}_{{\bf{k}}}.
\label{Equation_30}
\end{eqnarray}
In the next section we will derive the self-consistent equations for the excitonic order parameter $\Delta$ and chemical potential using the form of the partition function given in Eq.(\ref{Equation_19}) and the form of the action of the system obtained in Eq.(\ref{Equation_28}).
% 
%========================================================================
\subsection{\label{sec:Section_4_2} Self-consistent equations}
%========================================================================
%
As we mentioned at the beginning of the Section \ref{sec:Section_4}, the EI state is a sum of local on-site electron-hole correlations (see in Fig.~\ref{fig:Fig_4}). The expectation value, given in the expression of the local EI order parameter in Eq.(\ref{Equation_18}), could be calculated in the frame of the path integral method \cite{Negele} as well as, the fermion density averages $n_{\rm x}=\left\langle\bar{\rm x}({\bf{r}}\tau){\rm x}({\bf{r}}\tau)\right\rangle$. 

We get a set of self-consistent equations for the EI order parameter $\Delta$, single-particle fermion densities $n_{b}$, $n_{a}$  and the EI chemical potential $\mu$ 
\begin{eqnarray}
&&\frac{1}{N}\sum_{{\bf{k}}}\left[f(E^{+}_{{\bf{k}}})+f(E^{-}_{{\bf{k}}})\right]=1,
\label{Equation_31} 
\newline\\
&&\tilde{n}=\frac{1}{N}\sum_{{\bf{k}}}\xi_{{\bf{k}}}\frac{f(E^{+}_{{\bf{k}}})-f(E^{-}_{{\bf{k}}})}{\sqrt{\xi^{2}_{{\bf{k}}}+4\Delta^{2}}},
\label{Equation_32} 
\newline\\
&&\Delta=-\frac{U\Delta}{N}\sum_{{\bf{k}}}\frac{f(E^{+}_{{\bf{k}}})-f(E^{-}_{{\bf{k}}})}{\sqrt{\xi^{2}_{{\bf{k}}}+4\Delta^{2}}}.
\label{Equation_33}  
\end{eqnarray}
Here, $\xi_{{\bf{k}}}=-{t}_{{\bf{k}}}+\bar{\epsilon}_{a}+\tilde{t}_{{\bf{k}}}-\bar{\epsilon}_{b}$ is the quasiparticle dispersion, and the energy parameters $E^{+}_{{\bf{k}}}$ and $E^{-}_{{\bf{k}}}$ are 
\begin{eqnarray}
&&E^{\pm}_{{\bf{k}}}=\frac{1}{2}\left(-{t}_{{\bf{k}}}+\bar{\epsilon}_{a}-\tilde{t}_{{\bf{k}}}+\bar{\epsilon}_{b}\pm{\sqrt{\xi^{2}_{{\bf{k}}}+4\Delta^{2}}}\right).
\label{Equation_34}
\end{eqnarray}
Next, $f(\epsilon)$ denotes the Fermi-Dirac distribution function $f(\epsilon)=1/\left(e^{\beta{\epsilon}}+1\right)$.
%
%======================================================================================
\begin{figure}
\includegraphics[width=200px,height=35px]{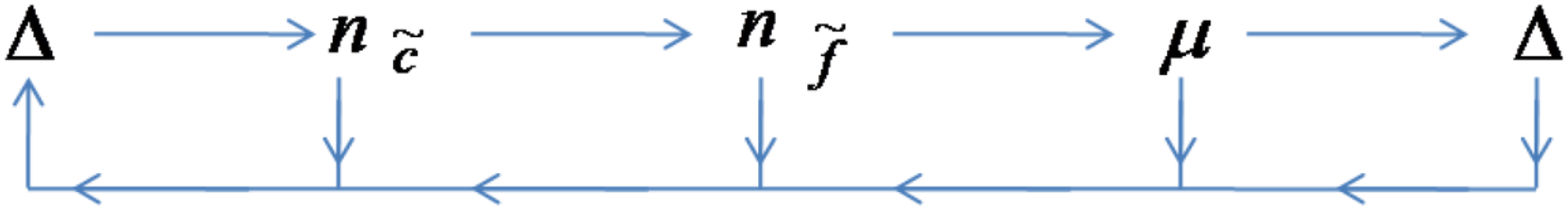}% Here is how to import EPS art
\caption{\label{fig:Fig_5}(Color online) The schematic diagram of the self-consistent solution procedure for the local EI order parameter $\Delta$.}
\end{figure}
%
%======================================================================================

The system of equations obtained in Eqs.(\ref{Equation_31})-(\ref{Equation_33}) is very analogue to the system of coupled equations obtained in Ref.\ \onlinecite{Seki}, where the authors employ the variational cluster approach (VCA) for the study of the EI state.  
In Fig.~\ref{fig:Fig_5} we have presented schematically the self-consistent cycle-procedure for calculating numerically the excitonic order parameter $\Delta$.  
%
%====================================================================
\subsection{\label{sec:Section_4_3} Numerical results}
%=====================================================================
%
The numerical solution of the system of self-consistent equations (\ref{Equation_31})-(\ref{Equation_33}) is performed for a fixed value of the total particle density $n=n_{\tilde{f}}+n_{\tilde{c}}=1$ and ${\bf{k}}$-summations were performed with the ($100$$\times$$100$) ${\bf{k}}$-points in the First Brillouin Zone (FBZ). The finite-difference approximation method is used, within the HYBRJ algorithm,\cite{Minpack} which retains the fast convergence of the Newton's method.\cite{Powel} The accuracy of convergence for numerical solutions is achieved with a relative error of order of $10^{-7}$ and a relatively small number of iterations was needed to get demanded convergence. 
%
%=================================================================================================
\begin{figure}    
 \centerline{\includegraphics[width=170px, height=450px]{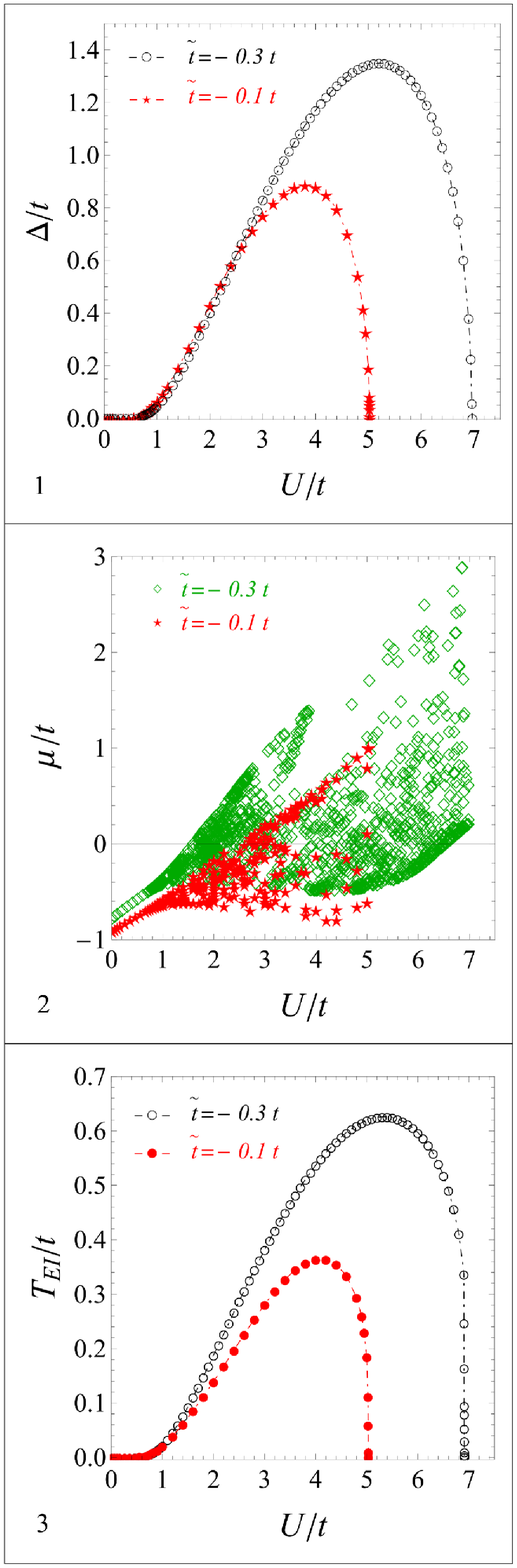}}% Here is how to import EPS art
\caption{\label{fig:Fig_6}}{(Color online) The numerical results for the excitonic order parameter $\Delta$ (top-panel-1), EI chemical potential (middle-panel-2) and critical temperature of the EI transition $T_{EI}$ (bottom-panel-3).}
\end{figure}
%=================================================================================================
%
The panel-1 in Fig.~\ref{fig:Fig_6} shows the numerical results for the local excitonic order parameter $\Delta$ for the EI state at zero temperature case. Two different values of $\tilde{t}$ are considered. 
The obtained values for the lower and upper
bounds of the Coulomb interaction parameter $U/t$ are about $(U_{c1}/t,U_{c2}/t)=(0.0,6.97)$ for the case $\tilde{t}=-0.3t$ and 
$(U_{c1}/t,U_{c2}/t)=(0.0,5.026)$ for the case $t_{f}=-0.1t$. They agree relatively well with the VCA results obtained in Ref.\onlinecite{Seki}.

The exact numerical solutions for the chemical potential
at $T = 0$ (thus, in the deep EI phase of the system) are plotted in the panel-2 in Fig.~\ref{fig:Fig_6}. Notably, for the intermediate and strong interaction limits ($0.78\leq U/t\leq 6.97$ for the case $\tilde{t}=-0.3t$) the chemical potential forms a
well defined band (see the leaf-like structures in the panel-2 in Fig.~\ref{fig:Fig_6}) and a single particle excitation gap
$\Delta_{H}\thicksim\mu_{\rm max}-\mu_{\rm min}$ is opening, indicating the SM-SC transition in the system of excitons.\cite{Seki} Here, $\mu_{max}$ and $\mu_{min}$ are
the upper and lower bounds of the chemical potential. 

We observe, that by moving from weak ($0.0\leq U/t\leq 0.78$ for $\tilde{t}=-0.3t$) into an intermediate coupling
regime ($0.78< U/t\leq 5.0$), the single-particle gap $\Delta_{H}$, and the pairing gap
parameter $\Delta$, both are increasing, while in the strong coupling limit ($5.0 < U/t\leq 6.97$ for $\tilde{t}=-0.3t$), $\Delta$
decreases rapidly with increasing the interaction parameter, and $\Delta_{H}$ remains
open (the Hartree-like gap structure).
Contrary, at very small values of the Coulomb interaction parameter, the chemical potential is 
coinciding with the values of the chemical potential at the EI transition line (see the discussion in Ref.\onlinecite{Apinyan_2}) (when $\Delta=0$ and $T\thicksim T_{EI}$), thus, collapsing into the single valued case $\mu \rightarrow\mu_{EI}$. In this case the single particle gap $\Delta_{H}$ also collapses: $\Delta_{H}\rightarrow 0$. It is also remarkable to note that, at the very small interaction region (SM side) the chemical is always negative $\mu<0$. Remember, that the binding energy of a molecule in the BEC limit is $E_{bind}\approx |2\bar{\mu}|$. \cite{Nozieres, Tsuchiya_1,Tsuchiya_2} 

The panel-3 in the bottom in Fig.~\ref{fig:Fig_6}, shows the exciton pair formation critical temperature dependence on the Coulomb interaction parameter $U/t$. The lines plotted there represent the beginning of the EI phase transition, i.e. when $\Delta(T_{EI},U)=0$. We observe a
very good agreement with the previously done mean-field, VCA and slave-boson (SB) results.\cite{Seki,Zenker_2,Zenker_3}

To evaluate the energy scales in the model we considering a $5$-nm GaAs coupled quantum well (QW), separated by a $4$-nm Al$_{0.33}$Ga$_{0.67}$As barrier \cite{Takahashi} with the effective electron mass m$_{e}=50.061 m_{0}$ (here $m_{0}$ is the free-electron mass) and in-plane effective hole-mass of around $0.1$m$_{0}$ (according with the Luttinger parameters \cite{Takahashi}), we use the exciton binding energy value $6.7$ meV, which corresponds to the $c$ electron hopping $t=4.96$ meV.\cite{Takahashi, Szymanska} For the energy scales corresponding to $\Delta$, we find for the quasi-2D GaAs/AlGaAs QW structure geometry $\Delta\approx 6.96 $ meV (corresponding to $U=5.2|t|=25.7$ meV), for the Hartree-gap $\Delta_{H}$ we should consider different interaction limits. In the BCS regime ($U=0.78|t|=0.156$ meV) we find $\Delta_{H}=0.019$ meV, in intermediate interaction regime ($U=3.84|t|=19$ meV) we have $\Delta_{H}=8.06$ meV and for the very strong coupling regime ($U=6.96|t|=34.5$ meV) we find $\Delta_{H}=13.84$ meV. For the energy scale of the excitonic pair formation critical temperature $T_{EI}$, we get $T_{EI}=3.1$ meV at $U=26.3$ meV, or, in temperature, units $T_{EI}=35.97$ K. 
%
%================================================================
\section{\label{sec:Section_5} Phase stiffness at $T=0$}
%================================================================
%
%======================================================================
\subsection{\label{sec:Section_5_1} Fermion integration}
%======================================================================
%
In this Section we will integrate out the fermions in the partition function given in Eq.(\ref{Equation_17}). This procedure is also described in the general diagram presented in Fig.~\ref{fig:Fig_1} and enables to obtain the  total  bosonic phase action of the system responsible for the phase stiffness mechanism. Especially, we would like to derive the phase stiffness parameter by showing how the nonlocal correlations between the preformed n.n. excitonic pairs are responsible for this. Then, we write the partition function in Eq.(\ref{Equation_17}) in the following form
\begin{eqnarray}
{\cal{Z}}=\int\left[{\cal{D}}\varphi\right] e^{-{\cal{S}}_{\rm eff}[\varphi]},
\label{Equation_35}
\end{eqnarray}
where the effective phase action is 
\begin{eqnarray}
{\cal{S}}_{\rm eff}[\varphi]=-\ln\int\left[{\cal{D}}\bar{b}{\cal{D}}b\right]\left[{\cal{D}}\bar{a}{\cal{D}}a\right] e^{-{{\cal{S}}}[\bar{a},a,\bar{b},b,\varphi]}.
\label{Equation_36}
\end{eqnarray}
We expand the logarithm up to second order in the cumulant series expansion and we find for the effective phase action
\begin{eqnarray}
{\cal{S}}_{\rm eff}[\varphi]=\tilde{{\cal{S}}}_{0}+\left\langle{{\cal{S}}}\right\rangle_{a,b}-\frac{1}{2}\left[\left\langle{\cal{S}}^{2}\right\rangle_{a,b}-\left\langle{\cal{S}}\right\rangle^{2}_{a,b}\right],
\label{Equation_37}
\end{eqnarray}
where $\tilde{{\cal{S}}}_{0}$ is an unimportant constant. Here, the statistical averages of the type $\left\langle ... \right\rangle_{a,b}$ are defined as usual \cite{Abrikosov}
\begin{eqnarray}
\left\langle...\right\rangle_{a,b}=\frac{\int\left[{\cal{D}}\bar{b}{\cal{D}}b\right]\left[{\cal{D}}\bar{a}{\cal{D}}a\right]...e^{-{\cal{S}}\left[\bar{a},a,\bar{b},b,\varphi\right]}}{\int\left[{\cal{D}}\bar{b}{\cal{D}}b\right]\left[{\cal{D}}\bar{a}{\cal{D}}a\right]e^{-{\cal{S}}\left[\bar{a},a,\bar{b},b,\varphi\right]}}.
\label{Equation_38}
\end{eqnarray}
The fermion averaging in Eq.(\ref{Equation_38}) can be considered as the quantum statistical averaging with the effective fermionic field action ${\cal{S}}_{\cal{F}}\left[\bar{a},a,\bar{b},b\right]$ obtained in Eq.(\ref{Equation_28}). This consideration is sometimes called like the Bogoliubov-mean-field self-consistency condition. Thus, we replace 
\begin{eqnarray}
\left\langle ... \right\rangle_{a,b}\rightarrow \left\langle ... \right\rangle_{{\cal{S}}_{\cal{F}}},
\label{Equation_39}
\end{eqnarray}
then, for the effective phase action we find 
\begin{eqnarray}
{\cal{S}}_{\rm eff}[\varphi]=\tilde{\cal{S}}_{0}+\left\langle{\cal{S}}\right\rangle_{{\cal{S}}_{\cal{F}}}-\frac{1}{2}\left[\left\langle{\cal{S}}^{2}\right\rangle_{{\cal{S}}_{\cal{F}}}-\left\langle{\cal{S}}\right\rangle^{2}_{{\cal{S}}_{\cal{F}}}\right].
\label{Equation_40}
\end{eqnarray}
After the self-consistency assumption in Eq.(\ref{Equation_39}), the average in Eq.(\ref{Equation_38}) could be rewritten as
\begin{eqnarray}
\left\langle ... \right\rangle_{{\cal{S}}_{\cal{F}}}=\frac{\int\left[{\cal{D}}\bar{b}{\cal{D}}b\right]\left[{\cal{D}}\bar{a}{\cal{D}}a\right]...e^{-{{\cal{S}}_{\cal{F}}\left[{\bar{a},a,\bar{b},b}\right]}}}{\int\left[{\cal{D}}\bar{b}{\cal{D}}b\right]\left[{\cal{D}}\bar{a}{\cal{D}}a\right]e^{-{{\cal{S}}_{\cal{F}}\left[{\bar{a},a,\bar{b},b}\right]}}}.
\label{Equation_41}
\end{eqnarray}
The relations in Eqs.(\ref{Equation_39}) and (\ref{Equation_41}) are principal for theory. Using them together, we resolve the problem of interacting fermions as a self-consistently coupled problem of the noninteracting fermions and the gauge-bosonic phase field. The bosonic sector plays the role of a suitable background (the glue), on which the collective excitations and correlations appear in the fermionic sector. We will examine the four-fermionic terms in Eq.(\ref{Equation_40}), the relevant part of nonlocal fermionic correlations. The important part of
the effective phase action given in Eq.(\ref{Equation_40}) is
\begin{eqnarray}
{\cal{S}}_{\rm eff}[\varphi]={\cal{S}}_{0}[\varphi]+{\cal{S}}_{J}[\varphi],
\label{Equation_42}
\end{eqnarray}
where the action ${\cal{S}}_{0}[\varphi]$ appears after considering the first order average term $\left\langle{\cal{S}}\right\rangle_{{\cal{S}}_{\cal{F}}}$: ${\cal{S}}_{0}[\varphi]=\left\langle{\cal{S}}\right\rangle_{{\cal{S}}_{\cal{F}}}$ and is given in Eq.(\ref{Equation_16}). The second term in Eq.(\ref{Equation_42}) is four-fermionic term
\begin{eqnarray}
{\cal{S}}_{J}[\varphi]=-\frac{1}{2}\left\langle{\cal{S}}^{2}\right\rangle_{{\cal{S}}_{\cal{F}}}.
\label{Equation_43}
\end{eqnarray}
We present here the evaluation of the second term given in Eq.(\ref{Equation_43}) and proportional to the product $t({\bf{r}}_{1},{\bf{r}}'_{1})\tilde{t}({\bf{r}}_{2},{\bf{r}}'_{2})$ (we kept formally
the lattice site notations in the hopping integrals for the
$b$ and $a$ bands). Derivation of the term proportional
to $\tilde{t}({\bf{r}}_{1},{\bf{r}}'_{1}){t}({\bf{r}}_{2},{\bf{r}}'_{2})$ is very similar. We have
\begin{eqnarray}
&&\frac{1}{2}\left\langle{\cal{S}}^{2}\right\rangle_{{\cal{S}}_{\cal{F}}}=\frac{1}{2}\sum_{\left\langle{\bf{r}}_{1},{\bf{r}}'_{1}\right\rangle}\sum_{\left\langle{\bf{r}}_{2},{\bf{r}}'_{2}\right\rangle}\int^{\beta}_{0}d\tau d\tau' \left[ t({\bf{r}}_{1},{\bf{r}}'_{1})\tilde{t}({\bf{r}}_{2},{\bf{r}}'_{2})
\left\langle \bar{a}({\bf{r}}_{1}\tau){a}({\bf{r}}'_{1}\tau)\bar{b}({\bf{r}}_{2}\tau')b({\bf{r}}'_{2}\tau')\right\rangle\times \right.
\nonumber\\
\nonumber\\
&&\left.
\times e^{-i\left[\varphi({\bf{r}}_{1}\tau)-\varphi({\bf{r}}'_{1}\tau)\right]}e^{-i\left[\varphi({\bf{r}}_{2}\tau')-\varphi({\bf{r}}'_{2}\tau')\right]}+t({\bf{r}}_{1},{\bf{r}}'_{1})\tilde{t}({\bf{r}}'_{2},{\bf{r}}_{2})\left\langle \bar{a}({\bf{r}}_{1}\tau)a({\bf{r}}'_{1}\tau)\bar{b}({\bf{r}}'_{2}\tau')b({\bf{r}}_{2}\tau')\right\rangle\times \right.
\nonumber\\
\nonumber\\
&&\left.
\times e^{-i\left[\varphi({\bf{r}}_{1}\tau)-\varphi({\bf{r}}'_{1}\tau)\right]}e^{i\left[\varphi({\bf{r}}_{2}\tau')-\varphi({\bf{r}}'_{2}\tau')\right]}+t({\bf{r}}'_{1},{\bf{r}}_{1})\tilde{t}({\bf{r}}_{2},{\bf{r}}'_{2})\left\langle \bar{a}({\bf{r}}'_{1}\tau)a({\bf{r}}_{1}\tau)\bar{b}({\bf{r}}_{2}\tau')b({\bf{r}}'_{2}\tau')\right\rangle\times \right.
\nonumber\\
\nonumber\\
&&\left.
\times e^{i\left[\varphi({\bf{r}}_{1}\tau)-\varphi({\bf{r}}'_{1}\tau)\right]}e^{-i\left[\varphi({\bf{r}}_{2}\tau')-\varphi({\bf{r}}'_{2}\tau')\right]}+t({\bf{r}}'_{1},{\bf{r}}_{1})\tilde{t}({\bf{r}}'_{2},{\bf{r}}_{2})\left\langle \bar{a}({\bf{r}}'_{1}\tau)a({\bf{r}}_{1}\tau)\bar{b}({\bf{r}}'_{2}\tau')b({\bf{r}}_{2}\tau')\right\rangle\times \right.
\nonumber\\
\nonumber\\
&&\left.
\times e^{i\left[\varphi({\bf{r}}_{1}\tau)-\varphi({\bf{r}}'_{1}\tau)\right]}e^{i\left[\varphi({\bf{r}}_{2}\tau')-\varphi({\bf{r}}'_{2}\tau')\right]}\right].
\label{Equation_44}
\end{eqnarray}
The action, in the form given in Eq.(\ref{Equation_44}), is the source for the nonlocal excitonic correlations in the considered system.
%
%======================================================================
\subsection{\label{sec:Section_5_2} The phase action}
%======================================================================
 
%
%=================================================================================================
\begin{figure}    
 \centerline{\includegraphics[width=180px, height=160px]{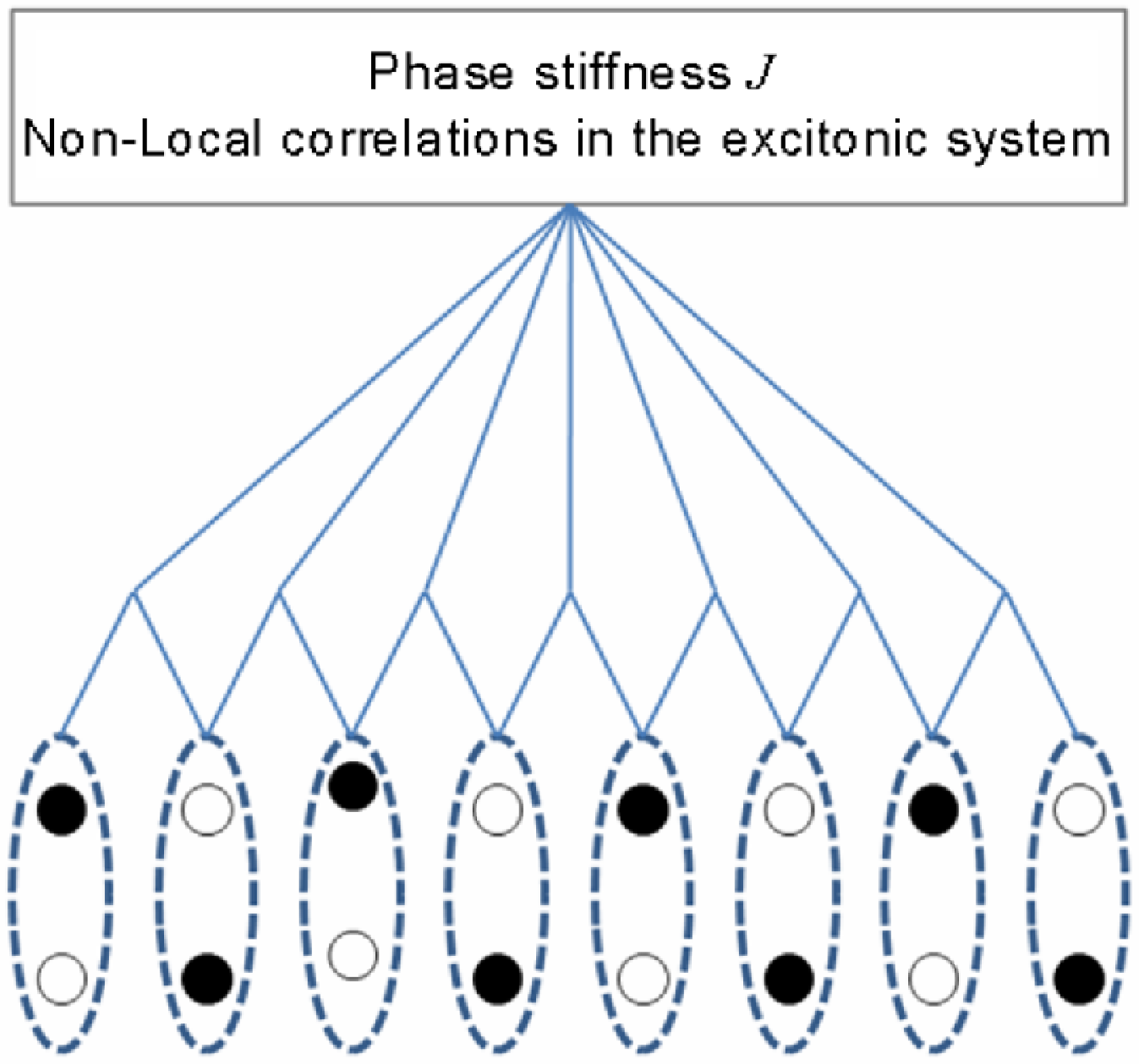}}% Here is how to import EPS art
\caption{\label{fig:Fig_7}}{(Color online) The nonlocal, exchange correlations, mediated by the fermion averaged exchange interactions between local (on-site) and nonlocal n.n. excitonic pairs, giving contributions to the total phase coherent state at $T=0$ (with the total phase stiffness parameter $J_{\rm}$).}
\end{figure}
%=================================================================================================

By dealing with the path integral formalism, we have the fermionic variables, which depend on the imaginary-time $\tau$. Thus, we use the many-body Wick averaging root for that case.\cite{Abrikosov} After calculating all averages in the right hand side in Eq.(\ref{Equation_44}), and after the not complicated evaluations, we rewrite the action ${\cal{S}}_{J}[\varphi]$ in the
form
\begin{widetext}
\begin{eqnarray}
{\cal{S}}_{J}\left[\varphi\right]=-{4t\tilde{t}}\int^{\beta}_{0}d\tau \int^{\beta}_{0}d\tau' \sum_{{\bf{r}}{\bf{r}}'}&&\left\{ \frac{}{} {\cal{F}}_{\rm ab}({\bf{r}}-{\bf{r}}',\tau-\tau'){\cal{F}}_{\rm ba}({\bf{r}}-{\bf{r}}',\tau'-\tau)\cos\left[\varphi({\bf{r}}\tau)+\varphi({\bf{r}}\tau')-\varphi({\bf{r}}'\tau)-\varphi({\bf{r}}'\tau')\right]\right.
\nonumber\\
&&\left.+{\cal{F}}_{\rm ab}({\bf{0}},\tau-\tau'){\cal{F}}_{\rm ba}({\bf{0}},\tau'-\tau)\cos\left[\varphi({\bf{r}}\tau)-\varphi({\bf{r}}\tau')-\varphi({\bf{r}}'\tau)+\varphi({\bf{r}}'\tau')\right]\frac{}{}\right\},
\label{Equation_45}
\end{eqnarray}
\end{widetext}
where ${\cal{F}}_{\rm ab}({\bf{r}}-{\bf{r}}',\tau-\tau')$ and ${\cal{F}}_{\rm ba}({\bf{r}}-{\bf{r}}',\tau'-\tau)$ are anomalous Green functions for the excitons. They are defined as
\begin{eqnarray}
{\cal{F}}_{\rm ab}({\bf{r}}\tau, {\bf{r}}'\tau')=\langle \bar{a}({\bf{r}}\tau)b({\bf{r}}'\tau')\rangle.
\label{Equation_46}
\end{eqnarray}
It is not difficult to find the product of the anomalous Green functions 
\begin{widetext}
\begin{eqnarray}
&&{\cal{F}}_{\rm ab}(\tau-\tau'){\cal{F}}_{\rm ba}(\tau'-\tau)
=\frac{4\Delta^{2}U^{2}}{z^{2}(\beta{N})^{2}}\sum_{{\bf{k}},\nu_{n}}
\sum_{{\bf{k}}',\nu'_{n}}\frac{\epsilon\left({{\bf{k}}}\right)\epsilon\left({{\bf{k}}}'\right)e^{-i(\nu'_{n}-\nu_{n})\delta}}{\left[E^{a}_{{\bf{k}}}(\nu_{n})E^{b}_{{\bf{k}}}(\nu_{n})-|\Delta|^{2}\right]\cdot\left[E^{a}_{{\bf{k}}'}(\nu_{n}')E^{b}_{{\bf{k}}'}(\nu_{n}')-|\Delta|^{2}\right]},
\label{Equation_47}
\end{eqnarray}
\end{widetext}
where $z$ is the number of n.n. sites ($z=4$ for the simple quadratic lattice). The phase action ${\cal{S}}_{J}\left[\varphi\right]$ in Eq.(\ref{Equation_45}) will take the form

\begin{eqnarray}
{\cal{S}}_{J}\left[\varphi\right]=-\frac{1}{2}\int^{\beta}_{0}d\tau \sum_{\left\langle {\bf{r}},{\bf{r}}'\right\rangle}J({\bf{r}}\tau,{\bf{r}}'\tau')\cos{2\left[\varphi({\bf{r}}\tau)-\varphi({\bf{r}}'\tau)\right]}.
\label{Equation_48} 
\end{eqnarray}
Here, $J({\bf{r}}\tau,{\bf{r}}'\tau')$ is the phase stiffness parameter and it is given by the relation
\begin{widetext}
\begin{eqnarray}
J=\frac{\Delta^{2}t\tilde{t}}{{N^{2}}}\sum_{{\bf{k}},{\bf{k}}'}\frac{\epsilon\left({{\bf{k}}}\right)\epsilon\left({{\bf{k}}}'\right)}{{\sqrt{\xi^{2}_{{\bf{k}}}+4\Delta^{2}}}}\left[\Lambda_{1}({\bf{k}},{\bf{k}}')\tanh\left(\frac{\beta E^{+}_{{\bf{k}}}}{2}\right)-\Lambda_{2}({\bf{k}},{\bf{k}}')\tanh\left(\frac{\beta E^{-}_{{\bf{k}}}}{2}\right)\right],
\label{Equation_49}
\end{eqnarray}
\end{widetext}
where we have already integrated  over the imaginary time $\tau'$ in Eq.(\ref{Equation_45}) and we have performed the Matsubara frequency summations in  Eq.(\ref{Equation_47}).
The parameters $\Lambda_{1}({\bf{k}},{\bf{k}}')$ and $\Lambda_{2}({\bf{k}},{\bf{k}}')$ in Eq.(\ref{Equation_49}) are given by
\begin{eqnarray}
\Lambda_{1}({\bf{k}},{\bf{k}}')=\frac{1}{E^{+}_{{\bf{k}}} - E^{+}_{{\bf{k}}'}}\cdot\frac{1}{E^{+}_{{\bf{k}}} - E^{-}_{{\bf{k}}'}},
\label{Equation_50}
\newline\\
\Lambda_{2}({\bf{k}},{\bf{k}}')=\frac{1}{E^{-}_{{\bf{k}}} - E^{-}_{{\bf{k}}'}}\cdot\frac{1}{E^{-}_{{\bf{k}}} - E^{+}_{{\bf{k}}'}}.
\label{Equation_51}
\end{eqnarray}
In the numerical evaluation of the phase coupling parameter $J$, given in Eq.(\ref{Equation_49}), it is pivotal to transform the ${\bf{k}}$-summations into energy integrals by introducing the tight-binding density of states (DOS) for the simple 2D quadratic lattice. \cite{Auerbach} The reason for this is encoded in the expression of denominators in Eqs.(\ref{Equation_50}) and (\ref{Equation_51}). In fact, the energy parameters $E^{+}_{{\bf{k}}}$ and $E^{+}_{{\bf{k}}'}$ (and $E^{-}_{{\bf{k}}}$, $E^{-}_{{\bf{k}}'}$) in denominators in Eqs.(\ref{Equation_50}) and (\ref{Equation_51}), are very close when ${\bf{k}}$ varies continuously, thus leading to a very strong divergent character of the parameters $\Lambda_{i}({\bf{k}},{\bf{k}}')$ ($i=1,2$). The use of density of states and the subsequent integration lead to the smoothing of those singularities, assuring the effective, finite solution for the excitonic phase stiffness parameter $J$. Thus, we introduce 2D elliptic DOS function $\rho_{2D}(x)$ for the simple quadratic lattice in the case of noninteracting regime
\begin{eqnarray}
\rho_{2D}(x)=\frac{1}{N}\sum_{{\bf{k}}}\delta\left[x-\epsilon\left({{\bf{k}}}\right)\right]=K\left(1-x^{2}/4\right)/\pi^{2},
\label{Equation_52}
\end{eqnarray}
where $K(x)$ stands for the complete elliptic integral of the first kind.\cite{Abramovich} Then, we rewrite Eq.(\ref{Equation_49}) in the integral form
\begin{widetext}
\begin{eqnarray}
J={\Delta^{2}t\tilde{t}}\int\int{dxdy}\frac{\rho_{2D}(x)\rho_{2D}(y)\epsilon\left(x\right)\epsilon\left(y\right)}{{\sqrt{\xi^{2}(x)+4\Delta^{2}}}}\left[\Lambda_{1}(x,y)\tanh\left(\frac{\beta E^{+}(x)}{2}\right)-\Lambda_{2}(x,y)\tanh\left(\frac{\beta E^{-}(x)}{2}\right)\right]
\label{Equation_53}
\end{eqnarray}
\end{widetext}
with the parameters $\Lambda_{1}(x,y)$ and $\Lambda_{2}(x,y)$ in Eq.(\ref{Equation_53}), being the continuous versions of the same parameters given in Eqs.(\ref{Equation_50}) and (\ref{Equation_51}).

\begin{figure}    
 \centerline{\includegraphics[width=210px, height=210px]{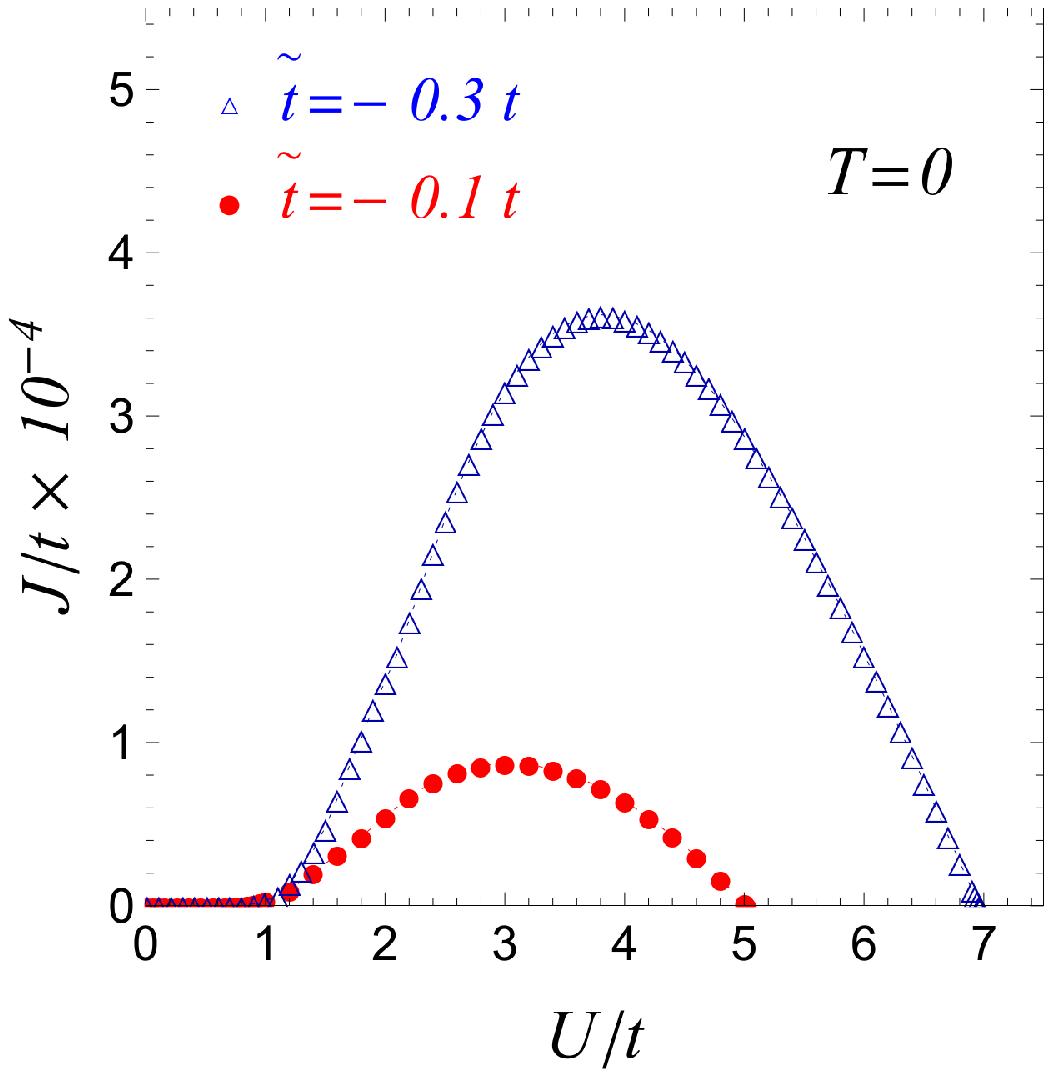}}% Here is how to import EPS art
\caption{\label{fig:Fig_8}}{(Color online) Numerical results for the excitonic phase stiffness parameter $J$ for 2D system and at $T=0$.}
\end{figure}

As we see, Eq.(\ref{Equation_53}) relates phase stiffness parameter $J$ with the local pairing order parameter $\Delta$. Thereby, the exciton pair formation (not the condensation) is a necessary prerequisite for the phase coupling between the n.n. excitonic pairs. We will see, that at low-temperatures, the macroscopic phase coherence of preformed excitonic pairs leads to the excitonic BEC transition in the system. The numerical evaluations of $J$ for the case $T=0$ K are shown in Fig.~\ref{fig:Fig_8}.

At the end of this Section we would like to emphasize on the analytical form of the phase coupling parameter $J$. Especially, it follows from Eq.(\ref{Equation_53}) that the macroscopic phase coherence in the system is characterized by an energy scale $J\thicksim (\Delta t_{e}t_{h})/({t_{e}+t_{h}})$ for all values of the Coulomb interaction parameter $U$, which is related to the motion of the center of mass of e-h composed quasiparticle,\cite{Tomio, Apinyan_2} because $(t_{e}t_{h})/(t_{e}+t_{h}) \approx (m_{e}+m_{h})^{-1}$. For the strong interaction limit, we converge with the hard core Boson model, with the kinetic energy proportional to ${\Delta}t_{e}{t}_{h}/U$ ($\Delta$ being the local excitonic order parameter). Thereby, we have shown that nonlocal correlations between the electrons and holes of different n.n. excitonic pairs, are relevant for the excitonic condensation. As it is discussed in Ref.(\onlinecite{Shi}), the luminescence line-shapes in the excitonic systems can be analyzed in terms of the spectral density function of the excitonic gas in interaction, which also determines the excitonic center-of-mass
distribution, related to the condensation in the low temperature limit.

For the energy scale corresponding to $J$, in the case of quasi-2D GaAs/AlGaAs QW structure geometry (see the Section \ref{sec:Section_4_2} for structural details) we find: $J\approx 0.001796 $ meV (corresponding to $U=0.0188$ eV) or, in temperature units $J\approx 20$ mK. 
%
%=================================================================================================
%=================================================================================================
%
%=================================================================================================
\section{\label{sec:Section_6} 2D Excitonic BEC at T=0}
%=================================================================================================
%
%=================================================================================================
\subsection{\label{sec:Section_6_1} Phase-phase propagator}
%=================================================================================================
%
In the discussion above, we have derived the effective phase-only action ${\cal{S}}_{\rm eff}\left[\varphi\right]={\cal{S}}_{0}\left[\varphi\right]+{\cal{S}}_{J}\left[\varphi\right]$.
In the following, we cast the ${\cal{S}}_{\rm eff}\left[\varphi\right]$ into the quantum rotor representation. To proceed, we replace the phase degrees of freedom with complex, unimodular field $z({\bf{r}}\tau)=e^{i\varphi({\bf{r}}\tau)}$ which satisfies the periodic boundary condition $z({\bf{r}}\beta)=z({\bf{r}}0)$. 
We introduce the new variables $z({\bf{r}}\tau)$ into the partition function in Eq.(\ref{Equation_35}) using the following identity
\begin{eqnarray}
\int\left[{\cal{D}}\bar{z}{\cal{D}}z\right] \delta\left(\frac{}{}\sum_{{\bf{r}}}|z({\bf{r}}\tau)|^{2}-N\frac{}{}\right)\delta\left(z-e^{i{\varphi}({\bf{r}}\tau)}\right)\delta\left(\bar{z}-e^{-i{\varphi}({\bf{r}}\tau)}\right)=1.
\label{Equation_54}
\end{eqnarray}
The inherent unimodular constraint on the complex variables $z({\bf{r}}\tau)$ ($|z({\bf{r}}\tau)|^{2}=1$) implies that on average the following condition holds 
\begin{eqnarray}
\frac{1}{N}\sum_{{\bf{r}}}|z({\bf{r}}\tau)|^{2}=1,
\label{Equation_55}
\end{eqnarray}
which forms a spherical constraint on a set of unimodular variables $z({\bf{r}}\tau)$. This can be resolved by introducing the Lagrange multiplier $\lambda$ resulting from the Laplace transform\cite{Lavrentiev} of the functional delta representation
\begin{eqnarray}
\delta\left(\frac{}{}\sum_{{\bf{r}}}|z({\bf{r}}\tau)|^{2}-N\frac{}{}\right)=\int^{+i\infty}_{-i\infty}\left[\frac{{\cal{D}}\lambda}{2\pi{i}}\right]e^{-i\int^{\beta}_{0}d\tau\sum_{{\bf{r}}} \lambda\left(\frac{}{}|z({\bf{r}}\tau)|^{2}-1\frac{}{}\right)}.
\label{Equation_56}
\end{eqnarray}
This adds a quadratic term (in the $z$-field) to the phase action ${\cal{S}}_{\rm eff}\left[\varphi\right]$. 

Next, we rewrite the action in Eq.(\ref{Equation_48}) in a more convenient form, using the half-angle  trigonometric transformation rule 

\begin{eqnarray}
\cos 2\left[{\varphi}({\bf{r}}\tau)-{\varphi}({\bf{r}}'\tau)\right]=2\cos^{2}\left[{\varphi}({\bf{r}}\tau)-{\varphi}({\bf{r}}'\tau)\right]-1.
\label{Equation_57}
\end{eqnarray}
Then, in terms of the complex variables $z({\bf{r}}\tau)$, the transformation in Eq.(\ref{Equation_57}) leads to a biquadratic form of the phase action in Eq.(\ref{Equation_48}). We have
\begin{eqnarray}
{\cal{S}}_{{{J}}}\left[\varphi\right]\rightarrow{\cal{S}}_{J}\left[\bar{z},z\right]=-\frac{J}{4}\int^{\beta}_{0}d\tau\sum_{\left\langle{\bf{r}},{\bf{r}}'\right\rangle}\left[\bar{z}({\bf{r}}\tau)z({\bf{r}}'\tau)+c.c.\right]^{2}.
\label{Equation_58}
\end{eqnarray}
After all, we can rewrite now the partition function in the form
\begin{eqnarray}
{\cal{Z}}=\int\left[{\cal{D}}\lambda\right]\left[{\cal{D}}\bar{z}{\cal{D}}z\right]\left[{\cal{D}}\varphi\right]&&e^{-{\cal{S}}_{0}\left[\varphi\right]}e^{\frac{J}{4}\int^{\beta}_{0}d\tau\sum_{\left\langle{\bf{r}},{\bf{r}}'\right\rangle}\left[\bar{z}\left({\bf{r}}\tau\right)z\left({\bf{r}}'\tau\right)+c.c.\right]^{2}}e^{i\int^{\beta}_{0}d\tau\sum_{{\bf{r}}}\lambda\left(|z({\bf{r}}\tau)|^{2}-1\right)}
\nonumber\\
&&\times \delta\left(z-e^{i{\varphi}({\bf{r}}\tau)}\right)\cdot\delta\left(\bar{z}-e^{-i{\varphi}({\bf{r}}\tau)}\right).
\label{Equation_59}
\end{eqnarray}
Furthermore, we linearize the action in Eq.(\ref{Equation_58}) and then we integrate out the phase variables in Eq.(\ref{Equation_59}) (for details of calculations see in Appendix \ref{Section_A}).
We obtain
\begin{eqnarray}
{\cal{S}}_{J}\left[\varphi\right]=-2J{\mathrm{g}_{B}}\int^{\beta}_{0}d\tau \sum_{\left\langle{\bf{r}},{\bf{r}}'\right\rangle}\cos{\left[\varphi({\bf{r}}\tau)-\varphi({\bf{r}}'\tau)\right]},
\label{Equation_60}
\end{eqnarray}
where $g_{B}$ stands for the bandwidth-renormalization factor, given explicitly in Eq.(\ref{Equation_25}). The partition function of the system takes the form
\begin{eqnarray}
{\cal{Z}}=\int\left[{\cal{D}}\lambda\right]\left[{\cal{D}}\bar{z}{\cal{D}}{z}\right]e^{-{\cal{S}}_{\lambda}[\bar{z},z]},
\label{Equation_61}
\end{eqnarray}
where the action ${\cal{S}}_{\lambda}[\bar{z},z]$ is
\begin{eqnarray}
{\cal{S}}_{\lambda}[\bar{z},z]&&=\frac{1}{\beta{N}}\sum_{{\bf{k}}\omega_{n}}\bar{z}({\bf{k}}\omega_{n}){\cal{G}}^{-1}_{z}({\bf{k}}\omega_{n})z({\bf{k}}\omega_{n}).
\label{Equation_62}
\end{eqnarray}
Here, we employed the Fourier transformation of $z$-variables $z({\bf{r}}\tau)=\frac{1}{\beta{N}}\sum_{{\bf{k}},\omega_{n}}z({\bf{k}}\omega_{n})e^{i{\bf{k}}\cdot{\bf{r}}-i\omega_{n}\tau}$ with $\omega_{n}$, being the Bose-Matsubara frequencies \cite{Abrikosov} $\omega_{n}=\frac{2\pi{n}}{\beta}$, with ($n=0,\pm 1,\pm 2,...$).
Furthermore, ${\cal{G}}^{-1}_{z}({\bf{k}}\omega_{n})$ is the Fourier transformed form of the inverse bosonic Green function and is given as
\begin{eqnarray}
{\cal{G}}^{-1}_{z}({\bf{k}}\omega_{n})=\gamma^{-1}(\omega_{n})-4g_{B}J\epsilon({\bf{k}})-\lambda.
\label{Equation_63}
\end{eqnarray}
The function $\gamma^{-1}(\omega_{n})$ in Eq.(\ref{Equation_63}) is the inverse of the Fourier transformation of two-point phase correlation function $\gamma({\bf{r}}\tau,{\bf{r}}'\tau')$
\begin{eqnarray}
\gamma({\bf{r}}\tau,{\bf{r}}'\tau')=\frac{1}{{\cal{Z}}_{0}}\int \left[{\cal{D}}{\varphi}\right]e^{-{\cal{S}}_{0}[{\varphi}]}e^{i\left[{\varphi}({\bf{r}}\tau)-{\varphi}({\bf{r}}'\tau')\right]},
\label{Equation_64}
\end{eqnarray}
where ${\cal{Z}}_{0}$ is the statistical sum of the noninteracting set of quantum rotators
\begin{eqnarray}
{\cal{Z}}_{0}=\int\left[{\cal{D}}\varphi \right] e^{-{\cal{S}}_{0}[\varphi]}.
\label{Equation_65}
\end{eqnarray}
The function $\epsilon({\bf{k}})$ is given in Eq.(\ref{Equation_26}) in the Section \ref{sec:Section_4_1}.
The calculation of the Fourier transform $\gamma(\omega_{n})$ of the function in Eq.(\ref{Equation_64}) is straightforward
\begin{eqnarray}
\gamma(\omega_{n})=\frac{8}{U{\cal{Z}}_{0}}\sum^{+\infty}_{{m}=-\infty}\frac{e^{-\frac{U\beta}{4}\left({{m}}-\frac{2\bar{\mu}}{U}\right)^{2}}}{1-16\left[\frac{i\omega_{n}}{U}-\frac{1}{2}\left({{m}}-\frac{2\bar{\mu}}{U}\right)\right]^{2}},
\nonumber\\
\ \ \
\label{Equation_66} 
\end{eqnarray}
where
\begin{eqnarray}
{\cal{Z}}_{0}=\sum^{+\infty}_{{m}=-\infty}e^{-\frac{U\beta}{4}\left({{m}}-\frac{2\bar{\mu}}{U}\right)^{2}}.
\label{Equation_67}
\end{eqnarray}
The summations in Eqs.(\ref{Equation_66}) and (\ref{Equation_67}) are over the winding numbers $m$ of the U(1) group.

Next, the phase-phase propagator $G_{z}({\bf{r}}\tau,{\bf{r}}'\tau')$ will be written as
\begin{eqnarray}
G_{z}({\bf{r}}\tau,{\bf{r}}'\tau')=\left\langle z({\bf{r}}\tau)\bar{z}({\bf{r}}'\tau')\right\rangle,
\nonumber\\
\label{Equation_68}
\end{eqnarray}
where the average is defined with the help of the action in Eq.(\ref{Equation_62})
\begin{eqnarray}
\langle ... \rangle\equiv\frac{\int\left[{\cal{D}}\bar{z}{\cal{D}}{z}\right][{\cal{D}}\lambda] ... e^{-{\cal{S}}_{\lambda}[\bar{z},z]}}{\int\left[{\cal{D}}\bar{z}{\cal{D}}{z}\right][{\cal{D}}\lambda]e^{-{\cal{S}}_{\lambda}[\bar{z},z]}}.
\label{Equation_69}
\end{eqnarray}
In the thermodynamic limit, when $N\rightarrow \infty$, the integration over $\lambda$-field in Eq.(\ref{Equation_61}) can be performed exactly using the saddle-point method\cite{Negele} $\left. \frac{\delta {\cal{S}}_{\lambda}[\bar{z},z]}{\delta \lambda}\right|_{\lambda=\lambda_{0}}=0$. Hence, the average in Eq.(\ref{Equation_69}) becomes
\begin{eqnarray}
\langle ... \rangle\equiv\frac{\int\left[{\cal{D}}\bar{z}{\cal{D}}{z}\right] ... e^{-{\cal{S}}_{\lambda_{0}}[\bar{z},z]}}{\int\left[{\cal{D}}\bar{z}{\cal{D}}{z}\right]e^{-{\cal{S}}_{\lambda_{0}}[\bar{z},z]}}.
\label{Equation_70}
\end{eqnarray}
In the general case, the local expression of the phase-phase correlation function, in Eq.(\ref{Equation_68}) is equal to unity, but, at $T=0$, this low breaks down, because we have to consider the symmetry breaking related to the phase transition in the bosonic sector. Thus, critically, we have the fluctuation form $z({\bf{r}}\tau)=\left\langle e^{i\varphi}({\bf{r}}\tau)\right\rangle+\tilde{z}({\bf{r}}\tau)$, and the unimodularity constraint is broken.
In the very low temperature limit, considering the BEC of excitons, we have the spontaneous breaking of local U(1) gauge-symmetry related to the phase field, leading to the non-vanishing expectation value of the $\left\langle e^{i\varphi}({\bf{r}}\tau)\right\rangle$. In order to demonstrate this, we separate the single-particle states ${\bf{k}}=0$ by using the Bogoliubov displacement operation (see for details in Refs.\ \onlinecite{Moskalenko} and \onlinecite{Simanek}). We write for the complex variables $z({\bf{k}},\omega_{n})$
\begin{eqnarray}
z({\bf{k}},\omega_{n})=\beta{N}\psi_{0}\delta_{{\bf{k}},0}\delta_{\omega_{n},0}+\tilde{z}({\bf{k}},\omega_{n})(1-\delta_{{\bf{k}},0})(1-\delta_{\omega_{n},0}),
\label{Equation_71}
\end{eqnarray}
where $\psi_{0}$ is the condensate transition amplitude $\psi_{0}=\left\langle {z}({\bf{k}},\omega_{n})\right\rangle$ of the bosonic field.\cite{Apinyan_2, Apinyan_3, Landau} Next, $\tilde{z}({\bf{k}},\omega_{n})$ is the on-condensate, or the excitation part of the effective Bose-field.\cite{Landau}
The Fourier transformation of the phase-phase propagator $G_{z}({\bf{r}}\tau,{\bf{r}}'\tau')$ in Eq.(\ref{Equation_68}) is 
\begin{eqnarray}
G_{z}({\bf{r}}\tau,{\bf{r}}'\tau')=\frac{1}{\beta{N}}\sum_{{\bf{k}},\omega_{n}}\left\langle z({\bf{k}},\omega_{n})\bar{z}({\bf{k}},\omega_{n})\right\rangle e^{-i{\bf{k}}\cdot{\bf{d}}+i\omega_{n}\delta}.
\label{Equation_72}
\end{eqnarray}

We consider the expectation value $\left\langle z({\bf{k}},\omega_{n})\bar{z}({\bf{k}},\omega_{n})\right\rangle$ in the one-exciton local limit, i.e. when ${\bf{d}}={\bf{r}}'-{\bf{r}}=0$ and $\delta=\tau'-\tau=0$ and we should draw the condensate part by applying the transformation given in Eq.(\ref{Equation_71}). Hence, we have the Fourier transform of the phase-phase propagator in the form
\begin{eqnarray}
&&G_{z}({\bf{k}},\omega_{n})=\frac{1}{\beta{N}}\left\langle z({\bf{k}},\omega_{n})\bar{z}({\bf{k}},\omega_{n})\right\rangle
=\beta{N}|\psi_{0}|^{2}\cdot\delta_{{\bf{k}},0}\delta_{\omega_{n},0}+\tilde G_{z}({\bf{k}},\omega_{n}).
\label{Equation_73}
\end{eqnarray}
Thereby, in Eq.(\ref{Equation_73}) we separate the coherent macroscopic state for the bosonic part of the interacting excitonic system, and the excitonic BEC is expected in the next. We will see in the Section \ref{sec:Section_7_1} how the excitonic propagator will be decomposed after applying the displacement operation given in Eq.(\ref{Equation_71}).
%
%===============================================================
\subsection{\label{sec:Section_6_2} Exciton condensate at $T=0$}
%===============================================================
%
We consider here the expectation value $\left\langle z({\bf{k}},\omega_{n})\bar{z}({\bf{k}},\omega_{n})\right\rangle$ again in the one-exciton local limit, i.e., when ${\bf{d}}={\bf{r}}-{\bf{r}}'=0$ and $\tau-\tau'=0$. In this case, it follows from Eq.(\ref{Equation_68}) 
that $G_{z}({\bf{r}}\tau,{\bf{r}}\tau)=1$. Substituting $G_{z}({\bf{k}},\omega_{n})$ from Eq.(\ref{Equation_73}) into the Eq.(\ref{Equation_72}), we get
\begin{eqnarray}
1-|\psi_{0}|^{2}=\frac{1}{\beta{N}}\sum_{\substack{{\bf{k}}\neq 0 \\ \omega_{n}\neq 0}}\tilde{G}_{z}({\bf{k}},\omega_{n}).
\label{Equation_74}
\end{eqnarray}
Here, $\tilde{G}_{z}({\bf{k}},\omega_{n})$ is related to the on-condensate exctitation part of the bosonic sector \cite{Landau}
\begin{eqnarray}
&&\tilde{G}_{z}({\bf{k}},\omega_{n})=\frac{1}{\beta{N}}\left\langle \tilde{z}({\bf{k}},\omega_{n})\bar{\tilde{z}}({\bf{k}},\omega_{n})\right\rangle.
\label{Equation_75}
\end{eqnarray}
At $T=0$, Eq.(\ref{Equation_74}) defines the excitonic BEC transition probability function $|\psi_{0}|^{2}$. 
At the temperatures different from zero, we have $\psi_{0}=0$, and the uniform static order parameter susceptibility diverges (see the Thouless criterion,\cite{Thouless} discussed in the Section II in Ref.\ \onlinecite{Tomio}), thus $G^{-1}_{z}({\bf{k}}={{0}},\omega_{n}=0)=0$, from which, we can derive the critical value of the Lagrange multiplier $\lambda=\lambda_{0}$
\begin{eqnarray}
\gamma^{-1}(\omega_{n}=0)-4g_{B}J\epsilon({\bf{0}})-\lambda_{0}=0.
\label{Equation_76}
\end{eqnarray}
Furthermore, we find 
\begin{eqnarray}
\lambda_{0}=\frac{U}{8}-\frac{2\bar{\mu}^{2}}{U}-4g_{B}J\epsilon({\bf{0}}).
\label{Equation_77}
\end{eqnarray}
Then, Eq.(\ref{Equation_74}) could be rewritten as
\begin{eqnarray}
1-|\psi_{0}|^{2}=\frac{1}{\beta{N}}\sum_{\substack{{\bf{k}}\neq 0 \\ \omega_{n}\neq 0}}\frac{1}{\gamma^{-1}(\omega_{n})-4g_{B}J\epsilon({\bf{k}})-\lambda_{0}}.
\label{Equation_78}
\end{eqnarray}
%
%======================================================================================
\begin{figure}
\centerline{\includegraphics[width=200px,height=200px]{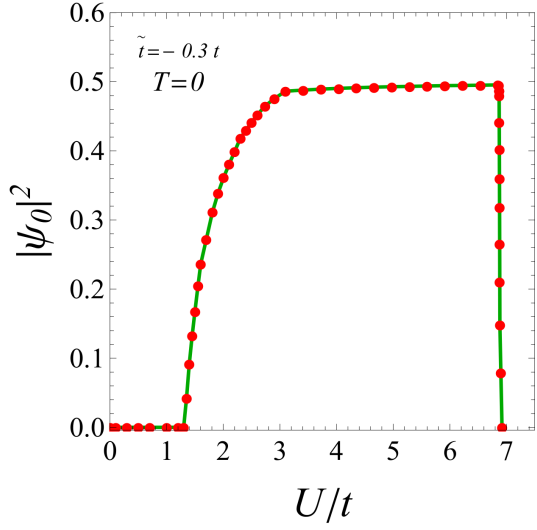}}% Here is how to import EPS art
\caption{\label{fig:Fig_9}(Color online) Exciton BEC transition probability, as a function of the Coulomb interaction parameter $U/t$. }
\end{figure}
%====================================================================

Furthermore, after Bose-Matsubara frequency summations in Eq.(\ref{Equation_78}), and in the limit, when $T\rightarrow 0$, we obtain the following equation for the BEC transition probability function 
\begin{eqnarray} |\psi_{0}|^{2}=1-\frac{U}{4N}\sum_{{\bf{k}}}\frac{1}{\sqrt{\bar{\mu}^{2}+2g_{B}U{J}\left[\epsilon({{\bf{0}}})-\epsilon({{\bf{k}}})\right]}}.
\label{Equation_79}
\end{eqnarray}
The values of $|\psi_{0}|^{2}$ are calculated and the results are plotted in Fig.~\ref{fig:Fig_9}, for the case $\tilde{t}=-0.3t$. We see that in the weak coupling regime, the transition probability function is negligibly small, i.e. in this case we have the BCS limit of the transition and this is consistent with the previous discussion. Contrary, for the intermediate values of the interaction parameter, the BEC transition probability function increases and remains practically constant for higher values of $U/t$ (for $U/t \gtrsim 3.0$ we have $|\psi_{0}|^{2}\approx 0.49$). In this case, we have the BEC limit of the phase transition. At the very strong limit of the coulomb interaction it starts to decrease drastically and disappears nearly for the same values that the gap parameter $\Delta$, critical temperature $T_{EI}$ and the phase stiffness parameter $J$. For a large domain of the Coulomb interaction parameter, the behavior of the function $|\psi_{0}|^{2}$ is very similar with the behavior of the Hartree gap $\Delta_{H}$ presented in the middle panel in Fig.~\ref{fig:Fig_6}, but the BEC amplitude disappears for the high values of the Coulomb interaction, while $\Delta_{H}$ does not. 
%
%============================================================================================
\section{\label{sec:Section_7} 2D Condensate DOS}
%=============================================================================================
%

%============================================================================================
\subsection{\label{sec:Section_7_1} Exciton spectral functions}
%=============================================================================================
%
The coherent hybridization between the valence band and conduction bands 
could be evidenced experimentally by examining the ARPES spectra, which measures the spectral intensities just above and below the temperature
$T_{EI}$ of excitonic pair formation. In ARPES experiments,
one observes the imaginary part of the real-time
retarded Green function, therefore the calculation of it
represents a remarkable importance. Our theoretical approach permits to access a variety of correlation functions in the system, which will give us  the information about the energy spectrum of the system. We have already defined the anomalous excitonic propagator ${\cal{F}}_{\rm ab}\left({\bf{r}}\tau,{\bf{r}};\tau'\right)$ in Eq.(\ref{Equation_46}), and the bosonic phase-phase propagator $G_{z}\left({\bf{r}}\tau,{\bf{r}};\tau'\right)$ in Eq.(\ref{Equation_68}). Here, we will also define the normal single-particle electronic Green functions for the $f$ and $c$ bands
\begin{eqnarray}
G_{\rm xx}({\bf{r}}\tau,{\bf{r}}'\tau')=-\langle {\rm x}({\bf{r}}\tau)\bar{\rm x}({\bf{r}}'\tau')\rangle.
\label{Equation_80}
\end{eqnarray}
After applying  the U(1) gauge transformation to the electrons, given in Eqs.(\ref{Equation_13}) and Eq.(\ref{Equation_14}), we will have for the normal and anomalous Green functions
\begin{eqnarray}
G_{\rm ff}({\bf{r}}\tau,{\bf{r}}'\tau')=\tilde{G}_{\rm bb}({\bf{r}}\tau,{\bf{r}}'\tau')\cdot\langle e^{-i\left[\varphi({\bf{r}}\tau)-\varphi({\bf{r}}'\tau')\right]}\rangle,
\label{Equation_81}
\newline\\
G_{\rm cc}({\bf{r}}\tau,{\bf{r}}'\tau')=\tilde{G}_{\rm aa}({\bf{r}}\tau,{\bf{r}}'\tau')\cdot\langle e^{-i\left[\varphi({\bf{r}}\tau)-\varphi({\bf{r}}'\tau')\right]}\rangle,
\label{Equation_82}
\end{eqnarray}
and 
\begin{eqnarray}
{\cal{F}}_{\rm cf}({\bf{r}}\tau,{\bf{r}}'\tau')={\tilde{\cal{F}}}_{\rm ab}({\bf{r}}\tau,{\bf{r}}'\tau')\cdot\langle e^{-i\left[\varphi({\bf{r}}\tau)-\varphi({\bf{r}}'\tau')\right]}\rangle,
\label{Equation_83}
\end{eqnarray}
where 
\begin{eqnarray}
\tilde{G}_{\rm bb}({\bf{r}}\tau,{\bf{r}}'\tau')=-\langle b({\bf{r}}\tau)\bar{b}({\bf{r}}'\tau'),
\label{Equation_84}
\newline\\
\tilde{G}_{\rm aa}({\bf{r}}\tau,{\bf{r}}'\tau')=-\langle a({\bf{r}}\tau)\bar{a}({\bf{r}}'\tau')\rangle,
\label{Equation_85}
\end{eqnarray}
and
\begin{eqnarray}
{\tilde{\cal{F}}}_{\rm ab}({\bf{r}}\tau,{\bf{r}}'\tau')=\langle \bar{a}({\bf{r}}\tau)b({\bf{r}}'\tau')\rangle
\label{Equation_86}
\end{eqnarray}
are new fermionic propagators, and the bosonic renormalization factor $G_{z}({\bf{r}}\tau, {\bf{r}}'\tau')$, in the expression of Green functions, is coinciding with the bandwidth renormalization factor defined in Eq.(\ref{Equation_25}) in the Section \ref{sec:Section_4}:
\begin{eqnarray}
G_{z}({\bf{r}}\tau, {\bf{r}}'\tau')=\langle e^{-i\left[\varphi({\bf{r}}\tau)-\varphi({\bf{r}}'\tau')\right]}\rangle.
\label{Equation_87}
\end{eqnarray}
For the U(1)-transformed Green functions, we write the Fourier transformation formula
\begin{eqnarray}
&&\tilde{G}_{\rm bb}({\bf{r}}\tau, {\bf{r}}'\tau')=\frac{1}{\beta{N}}\sum_{{\bf{k}},\nu_{n}}\tilde{G}_{\rm bb}({\bf{k}},\nu_{n}) e^{i{\bf{k}}\cdot({\bf{r}}-{\bf{r}}')-i\nu_{n}(\tau-\tau')},
\label{Equation_88}
\newline\\
&&\tilde{G}_{\rm aa}({\bf{r}}\tau, {\bf{r}}'\tau')=\frac{1}{\beta{N}}\sum_{{\bf{k}},\nu_{n}}\tilde{G}_{\rm aa}({\bf{k}},\nu_{n})e^{i{\bf{k}}\cdot({\bf{r}}-{\bf{r}}')-i\nu_{n}(\tau-\tau')},
\label{Equation_89}
\end{eqnarray}
and
\begin{eqnarray}
{\cal{F}}_{\rm cf}({\bf{r}}\tau, {\bf{r}}'\tau')=\frac{1}{\beta{N}}\sum_{{\bf{k}},\nu_{n}}{\tilde{\cal{F}}}_{\rm ab}({\bf{k}},\nu_{n})e^{i{\bf{k}}\cdot({\bf{r}}-{\bf{r}}')-i\nu_{n}(\tau-\tau')},
\label{Equation_90}
\end{eqnarray}
and for the bosonic propagator we have
\begin{eqnarray}
G_{z}({\bf{r}}\tau, {\bf{r}}'\tau')=\frac{1}{\beta{N}}\sum_{{\bf{k}},\omega_{n}}G_{z}({\bf{k}},\omega_{n})e^{i{\bf{k}}\cdot({\bf{r}}-{\bf{r}}')-i\omega_{n}(\tau-\tau')}.
\label{Equation_91}
\end{eqnarray}
Furthermore, for the Fourier transformations of the functions
in Eqs.(\ref{Equation_81})-(\ref{Equation_83}) we get the convoluted forms in the reciprocal ${\bf{k}}$-space
\begin{eqnarray}
G_{\rm ff}({\bf{k}},\nu_{n})=\frac{1}{\beta{N}}\sum_{{\bf{q}},\omega_{n}}G_{z}({\bf{q}},\omega_{n})\tilde{G}_{\rm bb}({\bf{k}}-{\bf{q}},\nu_{n}-\omega_{n}),
\label{Equation_92}
\newline\\
G_{\rm cc}({\bf{k}},\nu_{n})=\frac{1}{\beta{N}}\sum_{{\bf{q}},\omega_{n}}G_{z}({\bf{q}},\omega_{n})
\tilde{G}_{\rm aa}({\bf{k}}-{\bf{q}},\nu_{n}-\omega_{n}),
\label{Equation_93}
\end{eqnarray}
and 
\begin{eqnarray}
{\cal{F}}_{\rm cf}({\bf{k}},\nu_{n})=\frac{1}{\beta{N}}\sum_{{\bf{q}},\omega_{n}}G_{z}({\bf{q}},\omega_{n})
{\tilde{\cal{F}}}_{\rm ab}({\bf{k}}-{\bf{q}},\nu_{n}-\omega_{n}).
\label{Equation_94}
\end{eqnarray}
The summations in Eqs.(\ref{Equation_92})-(\ref{Equation_94}) are over the Bose-Matsubara frequencies $\omega_{n}={2\pi{n}/\beta}$.  
We will calculate the fermionic Green functions using the formalism discussed in the Sections \ref{sec:Section_2} and \ref{sec:Section_3} and also, functional derivation techniques.\cite{Negele} Then, for the $b$ ($f$) and $a$ ($c$)-band Green functions $\tilde{G}_{\rm x\rm {x}}({\bf{k}},\nu_{n})$, we get 
\begin{eqnarray}
\tilde{G}_{\rm bb}\left({\bf{k}},i\nu_{n}\right)=\frac{E^{a}_{{\bf{k}}}\left(\nu_{n}\right)}{E^{a}_{{\bf{k}}}\left(\nu_{n}\right) E^{b}_{{\bf{k}}}\left(\nu_{n}\right)-\Delta^{2}},
\label{Equation_95}
\newline\\
\tilde{G}_{\rm aa}\left({\bf{k}},i\nu_{n}\right)=\frac{E^{b}_{{\bf{k}}}\left(\nu_{n}\right)}{E^{a}_{{\bf{k}}}\left(\nu_{n}\right) E^{b}_{{\bf{k}}}\left(\nu_{n}\right)-\Delta^{2}}.
\label{Equation_96}
\end{eqnarray}
For the anomalous excitonic propagator we have 
\begin{eqnarray}
\tilde{{\cal{F}}}_{\rm ab}\left({\bf{k}},i\nu_{n}\right)=\frac{\Delta}{E^{a}_{{\bf{k}}}\left(\nu_{n}\right) E^{b}_{{\bf{k}}}\left(\nu_{n}\right)-\Delta^{2}}.
\label{Equation_97}
\end{eqnarray}
Next, we separate the condensate mode $\left\{ {\bf{q}}={{0}},\omega_{n}=0 \right\}$ in Eqs.(\ref{Equation_92}) and (\ref{Equation_94}). We have
\begin{eqnarray}
G_{\rm ff}({\bf{k}},\nu_{n})=|\psi_{0}|^{2} \cdot {\tilde{G}}_{\rm bb}({\bf{k}},\nu_{n})+\frac{1}{\beta{N}}\sum_{\substack{{\bf{q}}\neq 0 \\ \omega_{n}\neq 0}}\tilde G_{z}({\bf{q}},\omega_{n})\cdot \tilde{G}_{\rm bb}({\bf{k}}-{\bf{q}},\nu_{n}-\omega_{n}),
\label{Equation_98} 
\newline\\
G_{\rm cc}({\bf{k}},\nu_{n})=|\psi_{0}|^{2} \cdot {\tilde{G}}_{\rm aa}({\bf{k}},\nu_{n})+ \frac{1}{\beta{N}}\sum_{\substack{{\bf{q}}\neq 0 \\ \omega_{n}\neq 0}}\tilde G_{z}({\bf{q}},\omega_{n})\cdot \tilde{G}_{\rm aa}({\bf{k}}-{\bf{q}},\nu_{n}-\omega_{n}),
\label{Equation_99} 
\end{eqnarray}
and 
\begin{eqnarray}
{\cal{F}}_{\rm cf}({\bf{k}},\nu_{n})=|\psi_{0}|^{2} \cdot \tilde{\cal{F}}_{\rm ab}({\bf{k}},\nu_{n})
+\frac{1}{\beta{N}}\sum_{\substack{{\bf{q}}\neq 0 \\ \omega_{n}\neq 0}}\tilde G_{z}({\bf{q}},\omega_{n})\cdot \tilde{\cal{F}}_{\rm ab}({\bf{k}}-{\bf{q}},\nu_{n}-\omega_{n}).
\label{Equation_100}
\end{eqnarray}
As we see, the normal and excitonic propagators are composed of two parts, one, responsible for the condensate state, and the other for the on-condensate excitation part (see the discussion in Ref.\onlinecite{Landau} for the case of the degenerated nearly ideal Bose-gas). Note also, that first terms in the right-hand sides in Eqs.(\ref{Equation_98})-(\ref{Equation_100}) consist of the condensate-transition probability function $|\psi_{0}|^{2}$, multiplied with the fermionic propagators $\tilde{G}
_{\rm xx}({\bf{k}},\nu_{n})$ and $\tilde{F}_{\rm ab}({\bf{k}},\nu_{n})$. 

Now, we are ready to calculate the analytical forms of the normal excitonic spectral functions $A_{\rm xx}({\bf{k}},\nu)$ (${\rm x}=f,c$) and anomalous excitonic spectral function $A_{\rm cf}({\bf{k}},\nu)$ and, later on, the profiles of the DOS, including states participating to the condensate. We introduce here the spectral functions $A_{\rm xx}({\bf{k}},\nu)$ and $A_{\rm cf}({\bf{k}},\nu)$ that carries the same physical information as the correlation functions $G_{\rm xx}({\bf{k}},\nu_{n})$ and ${\cal{F}}_{\rm cf}({\bf{k}},\nu_{n})$. We define
\begin{eqnarray}
G_{\rm xx}({\bf{k}},\nu_{n})=\int^{+\infty}_{-\infty}d\nu'\frac{A_{\rm xx}({\bf{k}},\nu')}{i\nu_{n}-\nu'},
\label{Equation_101} 
\newline\\
{\cal{F}}_{\rm cf}({\bf{k}},\nu_{n})=\int^{+\infty}_{-\infty}d\nu'\frac{A_{\rm cf}({\bf{k}},\nu')}{i\nu_{n}-\nu'} .
\label{Equation_102}
\end{eqnarray}
The integrations here, are over continuous frequencies $\nu$. Note, that $G_{\rm xx}({\bf{k}},\nu_{n})$ and ${\cal{F}}_{\rm cf}({\bf{k}},\nu_{n})$ are total fermionic Green functions, which contain the bosonic sector. In the same way, we can introduce the spectral functions $A_{z}({\bf{k}},\nu)$, $\tilde{A}_{\rm bb}({\bf{k}},\nu)$, $\tilde{A}_{\rm aa}({\bf{k}},\nu)$ and $\tilde{A}_{\rm ab}({\bf{k}},\nu)$, associated with the charge and incoherent fermionic parts (without bosonic sector). They correspond respectively to the correlation functions $G_{z}({\bf{k}},\omega_{n})$, ${\tilde{G}}_{\rm bb}({\bf{k}},\nu_{n})$, ${\tilde{G}}_{\rm aa}({\bf{k}},\nu_{n})$ and ${\tilde{\cal{F}}}_{\rm ab}({\bf{k}},\nu_{n})$. We have the following equations for these counterparts
\begin{eqnarray}
G_{z}({\bf{k}},\omega_{n})=\int^{+\infty}_{-\infty}d\nu'\frac{A_{z}({\bf{k}},\nu')}{i\omega_{n}-\nu'},
\label{Equation_103} 
\newline\\
{\tilde{G}}_{\rm bb}({\bf{k}},\nu_{n})=\int^{+\infty}_{-\infty}d\nu'\frac{{\tilde{A}}_{\rm bb}({\bf{k}},\nu')}{i\nu_{n}-\nu'},
\label{Equation_104} 
\newline\\
{\tilde{G}}_{\rm aa}({\bf{k}},\nu_{n})=\int^{+\infty}_{-\infty}d\nu'\frac{{\tilde{A}}_{\rm aa}({\bf{k}},\nu')}{i\nu_{n}-\nu'},
\label{Equation_105} 
\end{eqnarray}
and 
\begin{eqnarray}
{\cal{F}}_{\rm ab}({\bf{k}},\nu_{n})=\int^{+\infty}_{-\infty}d\nu'\frac{A_{\rm ab}({\bf{k}},\nu')}{i\nu_{n}-\nu'}.
\label{Equation_106} 
\end{eqnarray}
Using these definitions, it is not difficult to show that the total spectral density functions $A_{\rm xx}({\bf{k}},\nu)$ and $A_{\rm cf}({\bf{k}},\nu)$ will take the forms
\begin{eqnarray}
&&A_{\rm ff}({\bf{k}},\nu)=|\psi_{0}|^{2}\cdot \tilde{A}_{\rm bb}({\bf{k}},\nu)-\frac{1}{N}\sum_{{\bf{q}}\neq 0}\int {d\nu'}A_{\rm z}({\bf{q}},\nu')\tilde{A}_{\rm bb}({\bf{k}}-{\bf{q}},\nu-\nu')\left[n(\nu')+f(\nu-\nu')\right],
\label{Equation_107}
\newline\\
&&A_{\rm cc}({\bf{k}},\nu)=|\psi_{0}|^{2}\cdot \tilde{A}_{\rm aa}({\bf{k}},\nu)-\frac{1}{N}\sum_{{\bf{q}}\neq 0}\int {d\nu'}A_{\rm z}({\bf{q}},\nu')\tilde{A}_{\rm aa}({\bf{k}}-{\bf{q}},\nu-\nu')\left[n(\nu')+f(\nu-\nu')\right],
\label{Equation_108}  
\end{eqnarray}
and 
\begin{eqnarray}
&&A_{\rm cf}({\bf{k}},\nu)=|\psi_{0}|^{2}\cdot \tilde{A}_{\rm ab}({\bf{k}},\nu)-\frac{1}{N}\sum_{{\bf{q}}\neq 0}\int {d\nu'}A_{\rm z}({\bf{q}},\nu')\tilde{A}_{\rm ab}({\bf{k}}-{\bf{q}},\nu-\nu')\left[n(\nu')+f(\nu-\nu')\right],
\label{Equation_109} 
\end{eqnarray}
where $n(\epsilon)=1/\left(e^{\beta\epsilon}-1\right)$ is the Bose-Einstein distribution function. 
Deriving the relations in Eqs.(\ref{Equation_107})-(\ref{Equation_109}), we have used the property of the function $n(\epsilon)$ of the complex argument: $n\left(i\nu_{n}+\epsilon\right)=-f\left(\epsilon\right)$.
%
%============================================================================================
\subsection{\label{sec:Section_7_2} DOS functions}
%=============================================================================================
%
The single-particle DOS is related with the imaginary part of the retarded Green function,\cite{Abrikosov} thus we need to
calculate the real-time retarded Green functions, which correspond to the Matsubara Green functions presented in the previous section. This could be done by the analytical continuation into the upper-half complex semi-plane ($\nu_{n}>0$) of fermionic frequency modes $i\nu_{n}$ 
\begin{eqnarray}
\tilde{G}^{\rm R}_{\rm bb}({\bf{k}},\nu)=\tilde{G}_{\rm bb}\left({\bf{k}},i\nu_{n}\right)|_{i\nu_{n}\rightarrow \nu+i\eta},
\label{Equation_110}
\newline\\
\tilde{G}^{\rm R}_{\rm aa}({\bf{k}},\nu)=\tilde{G}_{\rm aa}\left({\bf{k}},i\nu_{n}\right)|_{i\nu_{n}\rightarrow \nu+i\eta}.
\label{Equation_111}
\end{eqnarray}
The single-particle spectral functions are defined then as
\begin{eqnarray}
\tilde{\rho}_{\rm bb}\left({\bf{k}},\nu\right)=-\frac{1}{\pi}\operatorname{Im}\tilde{G}^{\rm R}_{\rm bb}({\bf{k}},\nu)=\left(\bar{\epsilon}_{\rm a}-\tilde{t}_{{\bf{k}}}-\nu\right)^{2}\delta\left[\left(\nu^{2}+A_{\bf{k}}\nu+B_{{\bf{k}}}\right)\cdot\left(\bar{\epsilon}_{\rm a}-\tilde{t}_{{\bf{k}}}-\nu\right)\right],
\nonumber\\
\label{Equation_112}
\newline\\
\tilde{\rho}_{\rm aa}\left({\bf{k}},\nu\right)=-\frac{1}{\pi}\operatorname{Im}\tilde{G}^{\rm R}_{\rm aa}({\bf{k}},\nu)=\left(\bar{\epsilon}_{\rm b}-\tilde{t}_{{\bf{k}}}-\nu\right)^{2}\delta\left[\left(\nu^{2}+A_{\bf{k}}\nu+B_{{\bf{k}}}\right)\cdot\left(\bar{\epsilon}_{\rm b}-\tilde{t}_{{\bf{k}}}-\nu\right)\right],
\nonumber\\
\label{Equation_113}
\end{eqnarray}
where 
\begin{eqnarray}
&&A_{\bf{k}}=t_{{\bf{k}}}+\tilde{t}_{\bf{k}}-\bar{\epsilon}_{\rm a}-\bar{\epsilon}_{\rm b}
\label{Equation_114}
\end{eqnarray}
and 
\begin{eqnarray}
B_{\bf{k}}=\bar{\epsilon}_{\rm a}\bar{\epsilon}_{\rm b}+4t\tilde{t}\epsilon^{2}({\bf{k}})-2\tilde{t}\bar{\epsilon}_{\rm a}\epsilon({\bf{k}})-2t\bar{\epsilon}_{\rm b}\epsilon({\bf{k}})-\Delta^{2}.
\label{Equation_115}
\end{eqnarray}
2D lattice dispersion $\epsilon({\bf{k}})$ in Eq.(\ref{Equation_115}) is given in Eq.(\ref{Equation_26}).
Next, ${\bf{k}}$-summed DOS will be 
\begin{eqnarray}
\tilde{\rho}_{\rm bb}\left(\nu\right)=\frac{1}{N}\sum_{\bf{k}}\tilde{\rho}_{\rm bb}\left({\bf{k}},\nu\right).
\label{Equation_116}
\newline\\
\tilde{\rho}_{\rm aa}\left(\nu\right)=\frac{1}{N}\sum_{\bf{k}}\tilde{\rho}_{\rm aa}\left({\bf{k}},\nu\right).
\label{Equation_117}
\end{eqnarray}

The summations over the wave vectors in Eqs.(\ref{Equation_116}) and (\ref{Equation_117}) can be simplified by introducing the appropriate DOS function for the 2D square lattice defined in the Section \ref{sec:Section_5_2} in Eq.(\ref{Equation_52}). Then, we have
\begin{widetext}
\begin{eqnarray} 
\tilde{\rho}_{\rm xx}\left(\nu\right)=\int^{+2.0}_{-2.0}dy \rho_{ 2D}(y)\frac{\left[\bar{\epsilon}_{{\rm \tilde{x}}}-\tilde{t}(y)-\nu\right]^{2}}{\sqrt{\xi^{2}(y)+4\Delta^{2}}}\cdot\left\{\frac{\delta\left[\nu-E^{+}(y)\right]}{|\bar{\epsilon}_{{\rm \tilde{x}}}-\tilde{t}(y)-E^{+}(y)|}+\frac{\delta\left[\nu-E^{-}(y)\right]}{|\bar{\epsilon}_{\rm \tilde{x}}-\tilde{t}(y)-E^{-}(y)|}\right\}.
\label{Equation_118}
\end{eqnarray}
\end{widetext}
Here, again $\tilde{t}(x)=2\tilde{t}x$  and the energy parameters $E^{\pm}(x)$, are continuous versions of parameters defined in Eq.(\ref{Equation_34}).

The retarded Green function, which corresponds to the anomalous function defined in Eq.(\ref{Equation_97}) is
\begin{eqnarray}
\tilde{{\cal{F}}}^{\rm R}_{\rm ab}({\bf{k}},\nu)=\tilde{{\cal{F}}}^{\rm R}_{\rm ab}\left({\bf{k}},i\nu_{n}\right)|_{i\nu_{n}\rightarrow \nu+i\eta} \ .
\label{Equation_119}
\end{eqnarray}
The single-particle excitonic spectral function corresponding to the anomalous Green function is then given by the relation
\begin{eqnarray}
\tilde{\rho}_{\rm ab}\left({\bf{k}},\nu\right)=-\frac{1}{\pi}\operatorname{Im} \tilde{{\cal{F}}}^{\rm R}_{\rm ab}({\bf{k}},\nu)=\Delta
\delta\left(\nu^{2}+A_{\bf{k}}\nu+B_{{\bf{k}}}\right).
\label{Equation_120}
\end{eqnarray}
The ${\bf{k}}$-summed anomalous DOS for the excitons will be
\begin{eqnarray}
\tilde{\rho}_{\rm ab}(\nu)=\Delta\left\{\frac{\rho_{2D}\left[\Lambda_{1}(\nu)\right]}{|\chi_{1}\left[\Lambda_{1}(\nu)\right]|}+\frac{\rho_{2D}\left[\Lambda_{2}(\nu)\right]}{|\chi_{2}\left[\Lambda_{2}(\nu)\right]|}\right\},
\label{Equation_121}
\end{eqnarray}
It has a simpler form than the function in the Eq.(\ref{Equation_118}).
The dimensionless parameters $\Lambda_{1,2}(\nu)$ are given by following expressions
\begin{widetext}
\begin{eqnarray}
\Lambda_{1}(\nu)=\frac{-\left[\left(t+\tilde{t}\right)\nu-\left(\bar{\epsilon}_{\tilde{c}}\tilde{t}+\bar{\epsilon}_{\tilde{f}}t\right)\right]+\sqrt{\left[\left(t-\tilde{t}\right)\nu+\left(\bar{\epsilon}_{\tilde{c}}\tilde{t}-\bar{\epsilon}_{\tilde{f}}t\right)\right]^{2}+4t\tilde{t}|\Delta|^{2}}}{4t\tilde{t}},
\label{Equation_122}
\newline\\
\Lambda_{2}(\nu)=\frac{-\left[\left(t+\tilde{t}\right)\nu-\left(\bar{\epsilon}_{\tilde{c}}\tilde{t}+\bar{\epsilon}_{\tilde{f}}t\right)\right]-\sqrt{\left[\left(t-\tilde{t}\right)\nu+\left(\bar{\epsilon}_{\tilde{c}}\tilde{t}-\bar{\epsilon}_{\tilde{f}}t\right)\right]^{2}+4t\tilde{t}|\Delta|^{2}}}{4t\tilde{t}}
\label{Equation_123}
\end{eqnarray}
\end{widetext}
and the functions $\chi_{i}\left[\Lambda_{1}(\nu)\right]$ ($i=1,2$), in the denominators in the right-hand side in Eq.(\ref{Equation_121}) are given as 

\begin{eqnarray}
\chi_{i}\left[\Lambda_{i}(\nu)\right]=2\left(t+\tilde{t}\right)\nu+8t\tilde{t}\Lambda_{i}(\nu)-2\left(\bar{\epsilon}_{a}\tilde{t}+\bar{\epsilon}_{b}t\right).
\label{Equation_124} 
\end{eqnarray}

Now, turning to the convolution forms of total fermionic and excitonic Green functions in Eqs.(\ref{Equation_98})-(\ref{Equation_100}), we need an explicit expression for the phase-bosonic Green function $G_{z}\left({\bf{k}},\omega_{n}\right)$. We will calculate it in the formalism of the effective phase action given in the quantum rotor model, discussed earlier in Section \ref{sec:Section_6} and also in Refs.\onlinecite{Apinyan_2} and \onlinecite{Apinyan_3}, where we have derived the effective phase-only action $S_{\rm eff}[\varphi]$ by integrating out the fermions. 

The retarded bosonic Green function \cite{Landau} is related to the Matsubara Green function, by the analytical continuation into the upper-half complex semi-plane ($\omega_{n}>0$) of complex bosonic frequencies $i\omega_{n}$
\begin{eqnarray}
G^{\rm R}_{z}({\bf{k}},\omega)=G_{z}({\bf{k}},i\omega_{n})|_{i\omega_{n}\rightarrow \omega+i\eta} 
\label{Equation_125}
\end{eqnarray}
and the ${\bf{k}}$-summed DOS for bosons reads as
\begin{eqnarray}
\rho_{z}(\omega)=-\frac{1}{\pi}\sum_{{\bf{k}}}\operatorname{Im}G^{\rm R}_{z}({\bf{k}},\omega).
\label{Equation_126}
\end{eqnarray}
After non difficult algebraic manipulations and replacing the summation in Eq.(\ref{Equation_79}) by integration with the help of 2D density of states, we have
\begin{eqnarray}
\rho_{z}(\omega)=-\frac{U}{4}\int^{+\infty}_{-\infty}dx \rho_{2D}\left(x\right)\left[\frac{\delta\left[\omega-\kappa_{1}(x)\right]}{\sqrt{\bar{\mu}^{2}+4UJ\left(2-x\right)}}+\frac{\delta\left[\omega-\kappa_{2}(x)\right]}{\sqrt{\bar{\mu}^{2}+4UJ\left(2-x\right)}}\right],
\label{Equation_127}
\end{eqnarray}
where $\kappa_{i}(x)$, $i=1,2$ are given by the following relations
\begin{eqnarray}
\kappa_{1,2}(x)=-\bar{\mu}\pm{\sqrt{\bar{\mu}^{2}+4U{J}\left(2-x\right)}}
\label{Equation_128}
\end{eqnarray}
and the stiffness parameter $J$ is given in Eq.(\ref{Equation_53}) in the Section \ref{sec:Section_5_2}. We see that the difference between $\kappa_{1}(x)$ and $\kappa_{2}(x)$ at the condensate mode $x=0$ gives exactly the binding energy of a molecule in the BEC limit: $E_{bind}\approx |2\bar{\mu}|$ (see also the discussion in the Section \ref{sec:Section_4_3}). As it could be expected, the bosonic DOS function Eq.(\ref{Equation_128}) is negative $\rho_{z}(\omega)< 0$. This is consistent with the general considerations of the weakly non-ideal Bose gas. \cite{Landau} 

Now, we are ready to calculate the coherent excitonic DOS functions for normal $f$ and $c$ bands and anomalous excitonic parts. We have
\begin{eqnarray}
&&\rho_{\rm ff}(\nu)=|\psi_{0}|^{2}\cdot \tilde{\rho}_{\rm bb}(\nu)
-U\int^{+2}_{-2}dx \frac{\rho_{2D}(x)}{4\sqrt{\bar{\mu}^{2}+4UJ\left(2-x\right)}}\times
\nonumber\\
&&\left\{\tilde{\rho}_{\rm bb}\left(\nu-\kappa_{1}\left(x\right)\right)\cdot\left[n\left(\kappa_{1}(x)\right)+f\left(\nu-\kappa_{1}(x)\right)\right]+\right.
\nonumber\\
&&\left.\tilde{\rho}_{\rm bb}\left(\nu-\kappa_{2}\left(x\right)\right)\cdot\left[n\left(\kappa_{2}(x)\right)+f\left(\nu-\kappa_{2}(x)\right)\right]\right\}
\label{Equation_129}
\end{eqnarray}
and
\begin{eqnarray}
&&\rho_{\rm cc}(\nu)=|\psi_{0}|^{2}\cdot \tilde{\rho}_{\rm aa}(\nu)
-U\int^{+2}_{-2}dx \frac{\rho_{2D}(x)}{4\sqrt{\bar{\mu}^{2}+4UJ\left(2-x\right)}}\times
\nonumber\\
&&\left\{\tilde{\rho}_{\rm aa}\left(\nu-\kappa_{1}\left(x\right)\right)\cdot\left[n\left(\kappa_{1}(x)\right)+f\left(\nu-\kappa_{1}(x)\right)\right]+\right.
\nonumber\\
&&\left.\tilde{\rho}_{\rm aa}\left(\nu-\kappa_{2}\left(x\right)\right)\cdot\left[n\left(\kappa_{2}(x)\right)+f\left(\nu-\kappa_{2}(x)\right)\right]\right\}.
\label{Equation_130}
\end{eqnarray}
For the anomalous excitonic DOS function, we have
\begin{eqnarray}
&&\rho_{\rm cf}(\nu)=|\psi_{0}|^{2}\cdot \tilde{\rho}_{\rm ab}(\nu)-U\int^{+2}_{-2}dx \frac{\rho_{2D}(x)}{4\sqrt{\bar{\mu}^{2}+4UJ\left(2-x\right)}}\times
\nonumber\\
&&\left\{\tilde{\rho}_{\rm ab}\left(\nu-\kappa_{1}\left(x\right)\right)\cdot\left[n\left(\kappa_{1}(x)\right)+f\left(\nu-\kappa_{1}(x)\right)\right]+\right.
\nonumber\\
&&\left.\tilde{\rho}_{\rm ab}\left(\nu-\kappa_{2}\left(x\right)\right)\cdot\left[n\left(\kappa_{2}(x)\right)+f\left(\nu-\kappa_{2}(x)\right)\right]\right\}.
\label{Equation_131}
\end{eqnarray}
The DOS functions given in Eqs.(\ref{Equation_129})-(\ref{Equation_131}) follow the same analytical structure as the excitonic spectral functions given in  Eqs.(\ref{Equation_107})-(\ref{Equation_109}). First terms, in the right hand sides in Eqs.(\ref{Equation_129})-(\ref{Equation_131}), represent the coherent condensate parts of the total DOS functions. As we will see later on, these parts for the normal excitonic DOS functions $\rho_{\rm xx}(\nu)$ follow the same behavior as the corresponding incoherent normal DOS functions (with an accuracy given by the factor $|\psi_{0}|^{2}$). \cite{Seki} The second terms in the right hand side in Eqs.(\ref{Equation_129})-(\ref{Equation_131}) are related to the on-condensate excitations in the system mediated by the strong bosonic field fluctuations. \cite{Landau} The sign $-$ near these terms comes from the sign of the pure bosonic DOS function given in Eq.(\ref{Equation_127}).
   
\begin{figure}
\includegraphics[width=270px, height=500px]{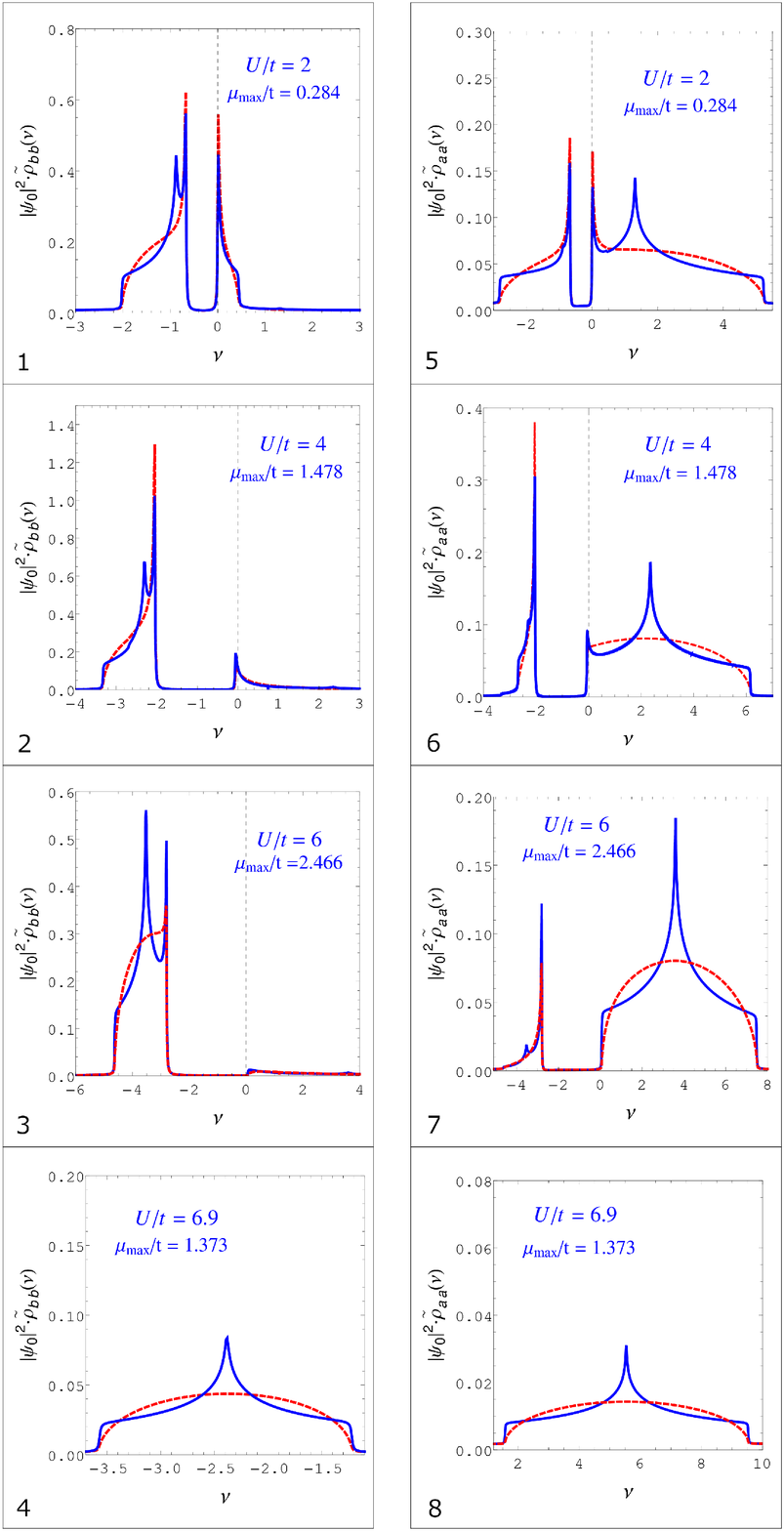}% Here is how to import EPS art
\caption{\label{fig:Fig_10}(Color online) Condensate DOS functions $|\psi_{0}|^{2}\tilde{\rho}_{\rm bb}$ and  $|\psi_{0}|^{2}\tilde{\rho}_{\rm aa}$ for different values of the Coulomb interaction parameter $U/t$ and for the case $T=0$. The case $\tilde{t}=-0.3t$ is considered here. The panels 1-4 or 5-8 in the figure show the SM-SC transition or the BCS-BEC crossover in the system of excitons.}
\end{figure}

\begin{figure}
\includegraphics[width=270px, height=400px]{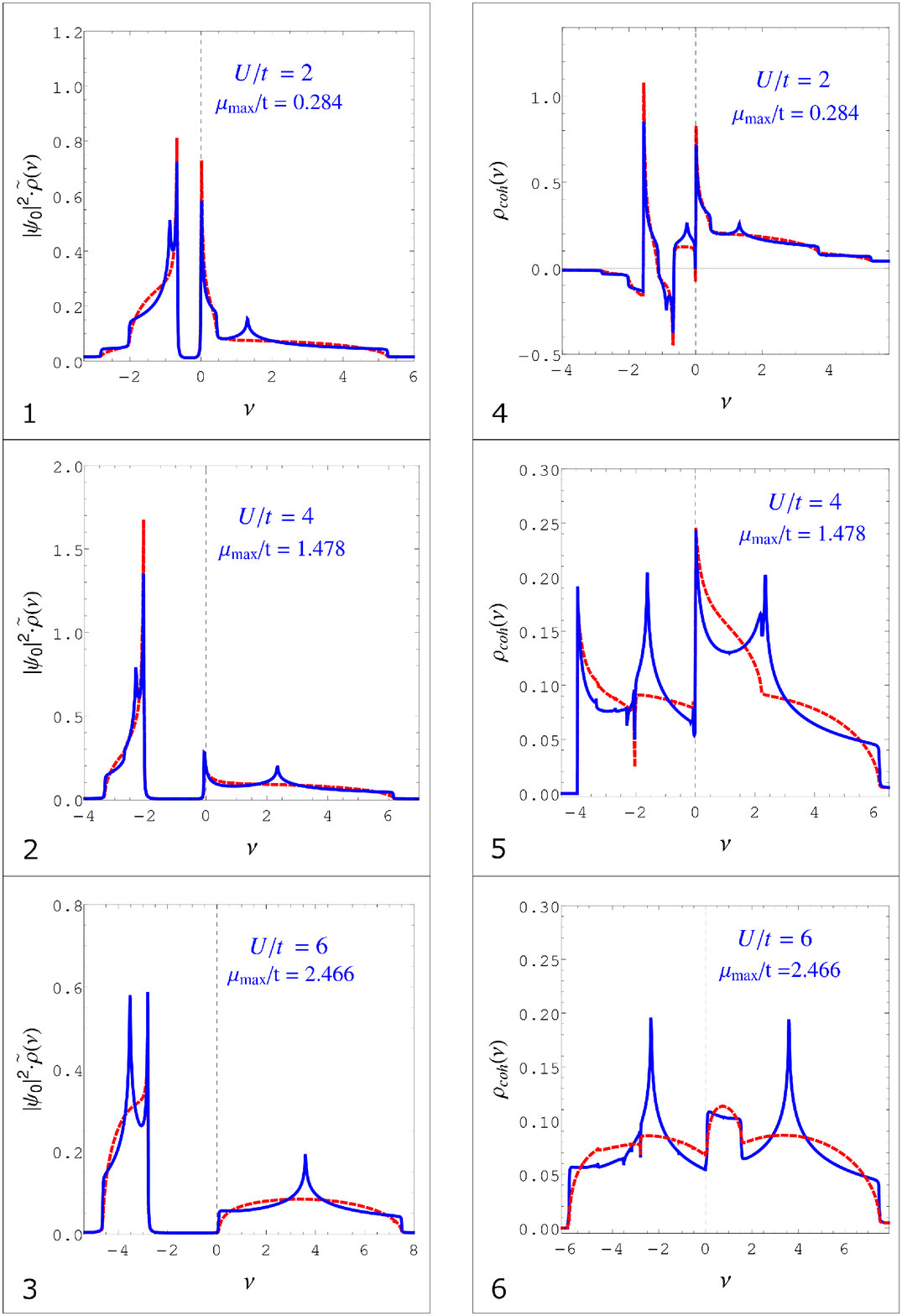}
\caption{\label{fig:Fig_11}(Color online) Total DOS $|\psi_{0}|^{2}\tilde{\rho}(\nu)$ (left panels) and total phase coherent DOS function $\rho_{\textit{\rm coh}}(\nu)$ (right panels) for different values of the Coulomb interaction parameter $U/t$ ($U/t=2$, $U/t=4$, $U/t=6$) and for the case $T=0$. The case $\tilde{t}=-0.3t$ is considered here.}
\end{figure}

The numerical evaluations of DOS functions at $T=0$ are given in Figs.~\ref{fig:Fig_10} and ~\ref{fig:Fig_11}, for the case $\tilde{t}=-0.3t$. The blue curves correspond to the Elliptic form of the noninteracting DOS (see in the Section \ref{sec:Section_5_2}). The dashed lines in red color, correspond to the semielliptic DOS structure, which corresponds to a Bethe lattice with an infinite number of nearest neighnors sites and we have
\begin{eqnarray}
\rho(x)=\sqrt{4-x^2}/(2\pi).
\label{Equation_132}
\end{eqnarray}
The experimental detection of the anomalous DOS function $\rho_{\rm cf}(\nu)$ is rather a complicated task, for this reason we present only the normal DOS function calculations for the excitonic system. During the numerical evaluations of the functions given in Eqs.(\ref{Equation_129})-(\ref{Equation_130}) we have used an adaptive 21-point integration routine (with an absolute error of order of $10^{-4}$ and with a relative error of order of $10^{-7}$) combined with the Wynn $\epsilon$-algorithm. \cite{Winalgorithm}
We examine the DOS behavior over the entire BCS-BEC
crossover region (i.e., for different values of the
Coulomb interaction $U$). An artificial Lorentzian broadening $\eta=0.01$ is used in numerical evaluations for the condensate DOS functions $\tilde{\rho}_{\rm xx}(\nu)$, and furthermore, for the total coherent DOS functions of $f$ and $c$ -
orbitals. Meanwhile, We take the upper-bound values $\mu_{max}$ of the chemical potential solutions (see in
the Section \ref{sec:Section_4_3}). The principal reason of it is that the BEC transition
amplitude $\psi_{0}$ has no physical solutions along the
lower-bound $\mu_{min}$ of the chemical potential. On the other hand, the values $\mu_{max}$ are most convenient, because they
minimize the Hamiltonian of the system (see in Eq.(\ref{Equation_1})).
In Fig.~\ref{fig:Fig_10} we have presented the results for partial DOS functions $|\psi_{0}|^{2}\tilde{\rho}_{\rm xx}(\nu)$ at $T=0$, given by the first terms in Eqs.(\ref{Equation_129}) and (\ref{Equation_130}), for $b$ and $a$ electrons. The results are well consistent with the MF calculations in Refs.\onlinecite{Seki} and \onlinecite{Czycholl}. In pictures, in Fig.~\ref{fig:Fig_10}, we keep the normalization factor $|\psi_{0}|^{2}$, in order to see also its influence on the partial DOS evaluation, or in other words, to follow the condensate parts of the total DOS functions in Eqs.(\ref{Equation_129}) and (\ref{Equation_130}), which correspond to the fundamental
state ${\bf{k}}=0$. Different values of the Coulomb interaction parameter $U$ are considered, starting from the BCS regime ($U/t=2$), passing the crossover regime ($U/t=4$) and the BEC regime ($U/t=6$ and $U/t=6.9$). We see in all pictures here, that a hybridization gap is always present in the DOS spetrum, for both, $b$ and $a$ bands. The principal reason of it is the non-vanishing Hartree-gap $\Delta_{H}$ in the energy spectrum discussed above, in the Section \ref{sec:Section_4_3}. The hybridization-gap is proportional to the interaction parameter $U$. We observe that the peaks in the DOS become more separated when increasing $U$. For the very high values of the Coulomb interaction, the peaked DOS structure disappears completely and the DOS becomes spread along the frequency axis (see the panels $4$ and $8$ in Fig.~\ref{fig:Fig_10}). 

Small values of the hybridization gap (for the case of the small coulomb interaction parameter $U$) in panels 1 and 5 demonstrate the fact that the system is in the SM (in the excitonic insulating regime) or BCS limit (in the phase stiffness regime) of the excitonic transition scenario. When augmenting the Coulomb interaction parameter (see the panels 2 and 6 and panels 3 and 7) the hybridization gap becomes larger, and in this case the system is continuously passing into the SC or the BEC limit of the transition. Thus the DOS functions plotted in Fig.~\ref{fig:Fig_10} describes either the SC-SM transition, or the BCS-BEC crossover mechanism in the system. We observe in the panels 4 and 8 that for the strong Coulomb interaction case, both bands show nearly the free particle DOS behavior with a single DOS peak. Thus, in this limit, we have practically free conduction band electrons and valence band holes and the EI state, or the excitonic condensate are completely absent. 

Meanwhile, the strong bosonic fluctuations effects are given in the right hand panels in Fig.~\ref{fig:Fig_11}, where the full, coherent DOS structures are shown.   

In Fig.~\ref{fig:Fig_11}, we give the numerical calculation results for the total DOS function $\tilde{\rho}(\nu)=\tilde{\rho}_{\rm aa}(\nu)+\tilde{\rho}_{\rm bb}(\nu)$ with a precision given by the factor $|\psi_{0}|^{2}$ (see the left panels in Fig.~\ref{fig:Fig_11}), and also the results for the total phase coherent DOS function ${\rho}_{\textit{\rm coh}}(\nu)={\rho}_{\rm ff}(\nu)+{\rho}_{\rm cc}(\nu)$ (see in Eqs.(\ref{Equation_129}) and (\ref{Equation_130}) and right panels 4-6 in Fig.~\ref{fig:Fig_11}).

Contrary to the purely condensate DOS structures given in Fig.~\ref{fig:Fig_10} and in the panels 1-3 in Fig.~\ref{fig:Fig_11}, the full coherent DOS function ${\rho}_{\rm coh}(\nu)$ shows completely different behavior. 
 We see in panels 4-6 that the hybridization gap is completely absent for all values of the Coulomb interaction parameter and the DOS amplitude is largely reduced. In the BCS limit ($U/t=2$), the phase coherent DOS function takes also the negative values. This is due to the strong, bosonic fluctuation effects for the very small Coulomb interaction values (see the panel $4$). For the medium and strong values of the Coulomb interaction parameter ($U/t=4,$ and $U/t=6$), DOS functions are positive for all frequency modes and the bosonic fluctuation effects are stabilizing (see the panels 5 and 6). We see, also, in panels $5$ and $6$, that the hybridization gap is disappearing and strong coherence effects reduce the DOS amplitude in comparison with the  behavior of the total condensate DOS $|\psi_{0}|^{2}\tilde{\rho}(\nu)$. The reason about this gapless DOS behavior in the medium and strong interaction limits is related to the phase coherence effects between $f$ and $c$ bands and which is due to the presence of the phase stiffness
mechanism considered here. In difference with the condensate DOS functions,
we have always a finite number
of states for all values of the frequency modes: $\rho_{\rm coh}(\nu)\neq 0$ (see the right panels in Fig.~\ref{fig:Fig_11})). We observe also that the bosonic phase coherence effects occur mainly in the region of the hybridization gap of the system without the phase stiffness, fulfilling it with a finite number of states smaller in amplitude than the excitonic formation peaks in the outermost sides of the hybridization region. 

The EI order parameter is not identical with the single-particle gap, especially in the BEC regime,
thus the experimental evidence of the EI state or the BEC of excitons could be
signaled only as the spontaneous hybridization between the valence
and conduction bands (remember that in the coherent excitonic phase stiffness regime the normal DOS behavior is the same as in the case of the incoherent situation, and only the DOS amplitudes are affected). Therefore, as the experimentally accessible quantities, we have the partial DOS functions $\tilde{\rho}_{\rm xx}(\nu)$. In fact, the direct experimental measurement of the excitonic density of states is rather a difficult task, which asks highly refined experimental techniques. We suggest that the condensate parts of the DOS functions given by the first terms in the right-hand sides of Eqs.(\ref{Equation_129}) and (\ref{Equation_130}) could be experimentally verified by analyzing of the transmittance and photoluminescence excitation (PLE) spectra. \cite{Klingshirn} Using the optics relations in the case of the quasi -2D thin film on a transparent substrate, one can obtain the absorption spectrum from the measured transmittance one. Furthermore, the analysis of the absorption spectra will provide a way to determine the partial excitonic DOS. Here, a special attention has to be paid when preparing the sample for the measurement. For the DOS spectra modeling procedure, the fitting parameters have to be taken into account, such, like the width of the nanocrystal size distribution (which is usually taken as Gaussian), the broadening parameter, and the thickness of the film.\cite{Vasilevskii} Let mention also that it is very difficult to extract the exciton density of states from only the photoluminescence (PL) line shapes (with a fixed photon energy), since the origin of the Stokes shift and PL line broadening.\cite{Efros} The measurement by the PL technique of the real absorption for the partial DOS is thus problematic, because of the scattering and further effects.    
In the conventional PLE technique, the fixed photon energy is changed into the excitation energy and PLE spectroscopy could be an alternative solution of that problem, assuming that the photoexcited e-h pairs always end-up at the lowest energy states. The spectrum measured by the PLE techniques is strongly correlated to the absorption spectrum of the sample. 

Meanwhile, in order to measure the total coherent excitonic density of states given in Eqs.(\ref{Equation_129}) and (\ref{Equation_130}) we suggest the sensible techniques of integrated photoluminescence excitation  (IPLE). The direct measurement of the coherent DOS by PLE techniques would be difficult in this case, due to the thickness of the sample and significant extinction caused by the film inhomogeneities. The IPLE spectrum, arising from the integration of the excitonic PL lineshape, gives a very good estimate for the shapes of the absorption spectrum, taking into account also the absorption spectra for higher energies.  
%
%======================================================================================================
%=======================================================================================================
%
%======================================================================================================
%=======================================================================================================

%===================================================
\section{\label{sec:Section_8} Summary and outlook}
%===================================================
%
We considered the problem of excitonic condensation in the 2D solid state system. The spinless extended Falicov-Kimball model was applied within the functional path integral formalism. We have shown that the electron operator factorization gives the possibility to handle the problem of excitonic phase transitions, in the strongly correlated electron systems, in a systematic self-consistent manner. We have shown that there exists an excitonic insulator phase transition, mediated by the local Coulomb interaction between the two solid state bands. We have calculated the local excitonic order parameter $\Delta$, and the critical temperature $T_{EI}$ of the excitonic pair formation, thus describing the stability region of the insulator state in the excitonic system. In the midst of calculations, we give the energy scale evaluations of all important physical parameters entering into the system.

Furthermore, we integrate out the fermionic variables and we consider the nonlocal nearest neighbors excitonic correlations. We have shown that there is a finite phase stiffness parameter between the pairs, which manage the straightness of the phase-phase correlations in the system. We have calculated it in the quantum rotor description and we evaluate the phase stiffness parameter numerically. Considering the Bogoliubov displacement operation in the bosonic sector of our complex formalism, we have separated the excitonic condensate mode from the on-condensate excitations. As a consequence, we have obtained the analytical forms of the excitonic spectral functions and density of states. The  probability function of the exciton condensate transition has been calculated numerically for the hopping parameter $\tilde{t}=-0.3t$ and at zero temperature. We have shown that the physical solutions for the condensate transition amplitude are given by the maximal values of the upper bound of the chemical potential obtained in the excitonic insulator phase. 

We have calculated the normal DOS functions for the $f$ and $c$ bands by considering the elliptic and Bethe type (with an infinite coordination number) of the noninteracting DOS structures. Meanwhile, The excitonic DOS function is evaluated analytically, and physical meaning of the obtained formula is explored. We have shown that there is a finite hybridization gap in the condensate normal DOS spectra, which is proportional to the Coulomb interaction parameter $U$. 
We have shown that for the small values of $U$ the hybridization gap is very small, thus the system is in the SM limit (or the BCS limit for the phase stiffness regime). For the intermediate and strong values of the Coulomb interaction parameter, the hybridization gap increases and the system passes into the SC state (or the BEC limit for the phase stiffness regime). Thus, from the DOS structure we prove the existence of the SM-SC transition (or BCS-BEC crossover for the phase stiffness regime) in the pure 2D excitonic system. 

Furthermore, we evaluated numerically the DOS functions in the regime of the phase stiffness (which we call also the condensate regime). Notice, that the phase coherent DOS is different in our case from the DOS given in the sense of the bands hybridization discussed above, and by the phase coherence we do not understand here just a possible direct hybridization between the $f$ and $c$ bands, but rather a phase stiffness mechanism governed by the nonlocal excitonic pair correlations. We have shown that for the total phase coherent DOS structures the hybridization gap is disappearing totally and the DOS structure shows a gapless behavior for all values of the Coulomb interaction $U/t$. The strong coherence effects, and strong bosonic field fluctuations fulfill the region of the hybridization gap and lead also to a significant decrease of the DOS amplitude for the case of the intermediate and strong values of the Coulomb interaction. 

For a quasi-2D GaAs/AlGaAs quantum well structure, we have evaluated numerically the relevant energy scales present in our model, such as the Hartree gap (for different regimes of the Coulomb interaction $U/t$), the excitonic pairing gap and the critical temperature of the excitonic insulator phase transition. The results agree quantitatively well the previous VCA and HFA results.\cite{Seki, Zenker_1, Zenker_2}  
Furthermore, we have also evaluated the excitonic phase stiffness parameter for the given heterostructure and we get a very small value for $J$: $J\approx$ 20 mK. 

Summarizing the obtained results and parallelizing the discussed formalism with the previous results, it is still very hard to answer the question, whether the excitonic condensation in the 2D system takes place at the same time with the development of the excitonic pair formation EI region. For a 3D case, we already know \cite{Apinyan_2, Apinyan_3} that this is not the case, and the excitonic condensate phase and the EI state are two different states of the matter, and the critical temperatures corresponding them are different. Recently, we have shown \cite{Apinyan_1} that at very low temperatures, the particle phase stiffness in the pure-2D excitonic system, governed by the nonlocal cross correlations, is responsible for the vortex-antivortex binding phase-field state, known as the Berezinskii-Kosterlitz-Thouless superfluid state, and the critical temperature of this transition is found very small.\cite{Apinyan_1} 

However, it is fundamental to clarify also, whether the excitonic BEC transition temperature is coinciding with the critical temperature of the excitonic superfluid phase transition, and for this, the quasi-2D system of excitons should be considered and the inter-layer exciton correlations should be properly included. We will consider this subject in the near future.

\widetext
\appendix
%
%==============================================================================
\section{\label{Section_A} The action ${\cal{S}}_{\lambda}\left[\bar{z},z\right]$}
%==============================================================================
%
The action in Eq.(\ref{Equation_58}) is quartic in unimodular $z$-field and could be decoupled with the help of the MF-like decoupling procedure
\begin{eqnarray}
\left[\bar{z}\left({\bf{r}}\tau\right)z\left({\bf{r}}'\tau\right)+c.c.\right]^{2}\rightarrow 4\left\langle{\bar{z}\left({\bf{r}}\tau\right)z\left({\bf{r}}'\tau\right)}\right\rangle\left[\bar{z}\left({\bf{r}}\tau\right)z\left({\bf{r}}'\tau\right)+c.c.\right]
\label{Equation_A1}
\end{eqnarray}
Then we get
\begin{eqnarray}
{\cal{S}}_{J}[\bar{z},z]=-2g_{B}J\int^{\beta}_{0}d\tau \sum_{\left\langle{\bf{r}},{\bf{r}}'\right\rangle}\bar{z}({\bf{r}}\tau)z({\bf{r}}'\tau).
\label{Equation_A2}
\end{eqnarray}
Now we will derive the action given in Eq.(\ref{Equation_61}). We start with the partition function given in Eq.(\ref{Equation_35}). We introduce the Fadeev-Popov resolution for the delta functions in Eq.(\ref{Equation_35}) by introducing the bosonic ghost-fields ${\eta}({\bf{r}}\tau)$ and ${\bar{\eta}}({\bf{r}}\tau)$ as
\begin{eqnarray}
\delta(z-e^{i\varphi\left({\bf{r}}\tau\right)})=\int\left[{\cal{D}}\bar{\eta}\right]e^{i\int^{\beta}_{0}d\tau\sum_{{\bf{r}}}\bar{\eta}({\bf{r}}\tau)\left(z-e^{i\varphi\left({\bf{r}}\tau\right)}\right)},
\label{Equation_A3}
\newline
\delta(\bar{z}-e^{-i\varphi\left({\bf{r}}\tau\right)})=\int\left[{\cal{D}}{\eta}\right]e^{i\int^{\beta}_{0}d\tau\sum_{{\bf{r}}}{\eta}({\bf{r}}\tau)\left(\bar{z}-e^{-i\varphi\left({\bf{r}}\tau\right)}\right)}.
\label{Equation_A4}
\end{eqnarray}
Then we write
\begin{eqnarray}
&&e^{i\int^{\beta}_{0}d\tau\sum_{{\bf{r}}}\bar{\eta}({\bf{r}}\tau)\left(z-e^{i\varphi\left({\bf{r}}\tau\right)}\right)}=\lim_{N\rightarrow \infty}\prod^{N}_{n=1}\prod_{{\bf{r}}}e^{i\epsilon_{n}\bar{\eta}({\bf{r}}\tau)\left(z-e^{i\varphi\left({\bf{r}}\tau\right)}\right)}=
\nonumber\\
&&=\lim_{N\rightarrow \infty}\prod^{N}_{n=1}\prod_{{\bf{r}}}e^{i\epsilon_{n}\bar{\eta}({\bf{r}}\tau)z\left({\bf{r}}\tau\right)}\left[1-i\epsilon_{n}\bar{\eta}({\bf{r}}\tau)e^{i\varphi({\bf{r}}\tau)}+\frac{1}{2!}\left(-i\epsilon_{n}\bar{\eta}({\bf{r}}\tau)e^{i\varphi({\bf{r}}\tau)}\right)^{2}+\frac{1}{3!}\left(-i\epsilon_{n}\bar{\eta}({\bf{r}}\tau)e^{i\varphi({\bf{r}}\tau)}\right)^{3}+...\right].
\nonumber\\
\label{Equation_A5}
\end{eqnarray}
We can write also an analogue expression for the exponential $e^{i\int^{\beta}_{0}d\tau\sum_{{\bf{r}}}{\eta}\left(\bar{z}-e^{-i\varphi\left({\bf{r}}\tau\right)}\right)}$. Thereby, we have
\begin{eqnarray}
&&e^{i\int^{\beta}_{0}d\tau\sum_{{\bf{r}}}\bar{\eta}\left(z-e^{i\varphi\left({\bf{r}}\tau\right)}\right)}e^{i\int^{\beta}_{0}d\tau\sum_{{\bf{r}}}{\eta}({\bf{r}}\tau)\left(\bar{z}-e^{-i\varphi\left({\bf{r}}\tau\right)}\right)}=\lim_{N\rightarrow \infty}\prod^{N}_{n,m=1}\prod_{{\bf{r}},{\bf{r}}'}\left[1+\left(-i\epsilon_{n}\bar{\eta}({\bf{r}}\tau)e^{i\varphi({\bf{r}}\tau)}\right)\left(-i\epsilon_{m}{\eta}({\bf{r}}'\tau')e^{-i\varphi({\bf{r}}'\tau')}\right)\right.
\nonumber\\
&&\left.+\left(\frac{1}{2!}\right)^{2}\left(-i\epsilon_{n}\bar{\eta}({\bf{r}}\tau)e^{i\varphi({\bf{r}}\tau)}\right)^{2}\left(-i\epsilon_{m}{\eta}({\bf{r}}'\tau')e^{-i\varphi({\bf{r}}'\tau')}\right)^{2}+...\right]e^{i\int^{\beta}_{0}d\tau\sum_{{\bf{r}}}\bar{\eta}({\bf{r}}\tau)z({\bf{r}}\tau)}e^{i\int^{\beta}_{0}d\tau\sum_{{\bf{r}}}z({\bf{r}}\tau){\eta}({\bf{r}}\tau)}.
\label{Equation_A6}
\end{eqnarray}
We put now the expression in Eq.(\ref{Equation_A6}) into the partition function in Eq.(\ref{Equation_59}) and we integrate out the phase variables $\varphi\left({\bf{r}}\tau\right)$
\begin{eqnarray}
{\cal{Z}}=\lim_{N\rightarrow \infty}\prod^{N}_{n,m=1}\prod_{{\bf{r}},{\bf{r}}'}\int\left[{\cal{D}}\lambda\right]\left[{\cal{D}}\bar{z}{\cal{D}}z\right]e^{2g_{B}J\int^{\beta}_{0}d\tau \sum_{\left\langle{\bf{r}},{\bf{r}}'\right\rangle}\bar{z}({\bf{r}}\tau)z({\bf{r}}'\tau)}e^{i\int^{\beta}_{0}d\tau\sum_{{\bf{r}}}\lambda\left(|z({\bf{r}}\tau)|^{2}-1\right)}\times
\nonumber\\
\times\left[1-\frac{1}{1!}{\bar{\eta}}({\bf{r}}\tau)\eta({\bf{r}}'\tau')\epsilon_{n}\epsilon_{m}\frac{\left\langle e^{i\left[\varphi({\bf{r}}\tau)-\varphi({\bf{r}}'\tau')\right]}\right\rangle}{1!}+\frac{1}{2!}{\bar{\eta}}^{2}({\bf{r}}\tau)\eta^{2}({\bf{r}}'\tau')\epsilon^{2}_{n}\epsilon^{2}_{m}\frac{\left\langle e^{i2\left[\varphi({\bf{r}}\tau)-\varphi({\bf{r}}'\tau')\right]}\right\rangle}{2!}-\right.
\nonumber\\
\left.-\frac{1}{3!}{\bar{\eta}}^{3}({\bf{r}}\tau)\eta^{3}({\bf{r}}'\tau')\epsilon^{3}_{n}\epsilon^{3}_{m}\frac{\left\langle e^{i3\left[\varphi({\bf{r}}\tau)-\varphi({\bf{r}}'\tau')\right]}\right\rangle}{3!}+...\right].
\label{Equation_A7}
\end{eqnarray}
The phase averages in Eq.(\ref{Equation_A7}) are given as 
\begin{eqnarray}
\left\langle e^{in\left[\varphi({\bf{r}}\tau)-\varphi({\bf{r}}'\tau')\right]}\right\rangle=\frac{\int\left[{\cal{D}}\varphi\right]e^{-{\cal{S}}_{0}\left[\varphi\right]}e^{in\left[\varphi({\bf{r}}\tau)-\varphi({\bf{r}}'\tau')\right]}}{\int\left[{\cal{D}}\varphi\right]e^{-{\cal{S}}_{0}\left[\varphi\right]}}.
\label{Equation_A8}
\end{eqnarray}
On the other hand, we can decouple the expression $\left\langle e^{in\left[\varphi({\bf{r}}\tau)-\varphi({\bf{r}}'\tau')\right]}\right\rangle$ using the MF like cumulant averaging procedure and we obtain $\left\langle e^{in\left[\varphi({\bf{r}}\tau)-\varphi({\bf{r}}'\tau')\right]}\right\rangle=\left\langle e^{i\left[\varphi({\bf{r}}\tau)-\varphi({\bf{r}}'\tau')\right]}\right\rangle n!$.
Then, we rewrite the expression in Eq.(\ref{Equation_A7}) in the simplest form
\begin{eqnarray}
{\cal{Z}}=\int\left[{\cal{D}}\lambda\right]\left[{\cal{D}}\bar{z}{\cal{D}}z\right]\left[{\cal{D}}\bar{\eta}{\cal{D}}\eta\right]e^{2g_{B}J\int^{\beta}_{0}d\tau \sum_{\left\langle{\bf{r}},{\bf{r}}'\right\rangle}\bar{z}({\bf{r}}\tau)z({\bf{r}}'\tau)}e^{i\int^{\beta}_{0}d\tau\sum_{{\bf{r}}}\lambda\left(|z({\bf{r}}\tau)|^{2}-1\right)}\times
\nonumber\\
\times e^{-\sum_{{\bf{r}},{\bf{r}}'}\int^{\beta}_{0}d\tau\int^{\beta}_{0}d\tau'\bar{\eta}\left({\bf{r}}\tau\right)\gamma\left({\bf{r}}\tau,{\bf{r}}'\tau'\right)\eta\left({\bf{r}}'\tau'\right)+i\int^{\beta}_{0}d\tau\sum_{{\bf{r}}}\bar{\eta}({\bf{r}}\tau)z({\bf{r}}\tau)+i\int^{\beta}_{0}d\tau\sum_{{\bf{r}}}\bar{z}({\bf{r}}\tau){\eta}({\bf{r}}\tau)},
\label{Equation_A9}
\end{eqnarray}
where we introduced the phase-phase correlation function $\gamma\left({\bf{r}}\tau,{\bf{r}}'\tau'\right)=\left\langle e^{i\left[\varphi({\bf{r}}\tau)-\varphi({\bf{r}}'\tau')\right]}\right\rangle$. Now, we integrate out the bosonic $\eta$-field, by employing the HS complex transformation for bosons
\begin{eqnarray}
\int{\frac{1}{N}\prod_{i}d\bar{\zeta}_{i}d\zeta{i}}e^{-\sum_{ij}\bar{\zeta}_{i}A^{-1}_{ij}\zeta_{j}+\sum_{i}[\bar{z}_{i}{\zeta}_{i}+z_{i}\bar{\zeta}_{i}]}=\left[\det{A}^{-1}\right]^{-1}e^{\sum_{ij}\bar{z}_{i}A_{ij}z_{j}},
\label{Equation_A10}
\end{eqnarray}
we get
\begin{eqnarray}
\int\left[{\cal{D}}\bar{\eta}{\cal{D}}\eta\right]e^{-\sum_{{\bf{r}},{\bf{r}}'}\int^{\beta}_{0}d\tau\int^{\beta}_{0}d\tau'\bar{\eta}\left({\bf{r}}\tau\right)\gamma\left({\bf{r}}\tau,{\bf{r}}'\tau'\right)\eta\left({\bf{r}}'\tau'\right)+i\int^{\beta}_{0}d\tau\sum_{{\bf{r}}}\bar{\eta}({\bf{r}}\tau)z({\bf{r}}\tau)+i\int^{\beta}_{0}d\tau\sum_{{\bf{r}}}\bar{z}({\bf{r}}\tau){\eta}({\bf{r}}\tau)}\approx
\nonumber\\
\approx e^{-\sum_{{\bf{r}},{\bf{r}}'}\int^{\beta}_{0}d\tau\int^{\beta}_{0}d\tau'\bar{z}\left({\bf{r}}\tau\right)\gamma^{-1}\left({\bf{r}}\tau,{\bf{r}}'\tau'\right)z\left({\bf{r}}'\tau'\right)}.
\label{Equation_A11}
\end{eqnarray}
For the partition function in Eq.(\ref{Equation_A9}) we have
\begin{eqnarray}
{\cal{Z}}=\int\left[{\cal{D}}\lambda\right]\left[{\cal{D}}\bar{z}{\cal{D}}z\right]e^{2g_{B}J\int^{\beta}_{0}d\tau \sum_{\left\langle{\bf{r}},{\bf{r}}'\right\rangle}\bar{z}({\bf{r}}\tau)z({\bf{r}}'\tau)}e^{i\int^{\beta}_{0}d\tau\sum_{{\bf{r}}}\lambda\left(|z({\bf{r}}\tau)|^{2}-1\right)} e^{-\sum_{{\bf{r}},{\bf{r}}'}\int^{\beta}_{0}d\tau\int^{\beta}_{0}d\tau'\bar{z}\left({\bf{r}}\tau\right)\gamma^{-1}\left({\bf{r}}\tau,{\bf{r}}'\tau'\right)z\left({\bf{r}}'\tau'\right)}
\label{Equation_A12}
\end{eqnarray}
or, similarly,
\begin{eqnarray}
{\cal{Z}}=\int\left[{\cal{D}}\lambda\right]\left[{\cal{D}}\bar{z}{\cal{D}}z\right]e^{-\sum_{{{\bf{r}}},{\bf{r}}'}\int^{\beta}_{0}d\tau\int^{\beta}_{0}d\tau'\bar{z}({\bf{r}}\tau){\cal{G}}^{-1}_{z}({\bf{r}}\tau,{\bf{r}}'\tau')z({\bf{r}}'\tau')},
\label{Equation_A13}
\end{eqnarray}
where ${\cal{G}}^{-1}_{z}({\bf{r}}\tau,{\bf{r}}'\tau')$ is the inverse of the real-space bosonic Green-function 
\begin{eqnarray}
{\cal{G}}^{-1}_{z}({\bf{r}}\tau,{\bf{r}}'\tau')=-2g_{B}J\delta(\tau-\tau')\delta({\bf{r}}-{\bf{r}}'-{\bf{d}})+\lambda\delta\left({{\bf{r}}-{\bf{r}}'}\right)\delta(\tau-\tau')+\gamma^{-1}({\bf{r}}\tau,{\bf{r}}'\tau').
\label{Equation_A14}
\end{eqnarray}
In fact, the phase-phase correlation function $\gamma\left({\bf{r}}\tau,{\bf{r}}'\tau'\right)$ has the form
\begin{eqnarray}
\gamma\left({\bf{r}}\tau,{\bf{r}}'\tau'\right)=\delta\left({{\bf{r}}-{\bf{r}}'}\right)e^{-\frac{U}{\beta}\sum^{\infty}_{n=1}\frac{1-\cos\left[\omega_{n}\left(\tau-\tau'\right)\right]}{\omega^{2}_{n}}} \sum_{\left\{m\right\}}e^{-\frac{U\beta}{4}\left[m({\bf{r}})-\frac{2\bar{\mu}}{U}\right]^{2}-\frac{U}{2}\left(m-\frac{2\bar{\mu}}{U}\right)\left(\tau-\tau'\right)},
\label{Equation_A15}
\end{eqnarray}
where $\left\{m\right\}$ forms an infinite set of U(1) winding numbers (see the Section \ref{sec:Section_2_2}). Transforming the $z$-variables into the Fourier space (see the Section \ref{sec:Section_6_1}), we can write the partition function in Eq.(\ref{Equation_A13}) as
\begin{eqnarray}
{\cal{Z}}=\int\left[{\cal{D}}\lambda\right]\left[{\cal{D}}\bar{z}{\cal{D}}z\right]e^{-\frac{1}{\beta{N}}\sum_{{\bf{k}},\omega_{n}}{\bar{z}}({\bf{k}}\omega_{n}){\cal{G}}^{-1}_{z}({\bf{k}},\omega_{n})z({\bf{k}}\omega_{n})}
\label{Equation_A16}
\end{eqnarray}
and now ${\cal{G}}^{-1}_{z}({\bf{k}},\omega_{n})$ is
\begin{eqnarray}
{\cal{G}}^{-1}_{z}({\bf{k}},\omega_{n})=\gamma^{-1}(\omega_{n})-4g_{B}J-\lambda,
\label{Equation_A17}
\end{eqnarray}
where $\gamma^{-1}(\omega_{n})$ is the inverse of the Fourier transformation $\gamma(\omega_{n})$ of $\gamma(\tau-\tau')$ given in Eq.(\ref{Equation_64}), in the Section \ref{sec:Section_6_1}.
%
%=================

%
\end{document}